\documentclass[a4paper,fleqn]{cas-sc}

\usepackage[authoryear]{natbib}

\usepackage{amssymb}
\usepackage{amsthm}
\usepackage{amsmath}
\usepackage{color}
\usepackage{multirow}
\usepackage{graphicx}
\usepackage{subcaption}
\usepackage{lineno}
\usepackage{nomencl}
\makenomenclature
\usepackage{hyperref}

\def\tsc#1{\csdef{#1}{\textsc{\lowercase{#1}}\xspace}}
\tsc{WGM}
\tsc{QE}

\begin{document}
\let\WriteBookmarks\relax
\def\floatpagepagefraction{1}
\def\textpagefraction{.001}

\shorttitle{Parametric study on the water impacting of a free-falling symmetric wedge}   

\shortauthors{Yujin Lu and Alessandro Del Buono}  

\title [mode = title]{Parametric study on the water impacting of a free-falling symmetric wedge based on the extended von Karman's momentum theory}



%

\author[inst1,inst2]{Yujin Lu}[type=author]





\credit{Conceptualization, Methodology, Software, Investigation, Data Curation, Visualization, Writing - Original Draft}


\affiliation[inst1]{organization={Nanjing University of Aeronautics and Astronautics},
            addressline={Yudao Street 29}, 
            city={Nanjing},
            postcode={210016}, 
            state={Jiangsu},
            country={People’s Republic of China}}
\affiliation[inst2]{organization={National Research Council–Institute of Marine Engineering (CNR-INM)},
            addressline={Via di Vallerano 139}, 
            city={Roma},
            postcode={00128}, 
            country={Italy}}

\author[inst2]{Alessandro {Del Buono}}[type=author]
\cormark[1]

\ead{alessandro.delbuono@inm.cnr.it}


\credit{Validation, Investigation, Formal analysis, Writing - Review \& Editing}

\author[inst1]{Tianhang Xiao}[type=author]
\cormark[2]
\ead{xthang@nuaa.edu.cn}
\credit{Conceptualization, Writing - Review \& Editing, Supervision}

\author[inst2]{Alessandro Iafrati}[type=author]
\credit{Project administration, Writing - Review \& Editing, Supervision}

\author[inst1]{Jinfa Xu}[type=author]
\credit{Resources, Supervision}

\author[inst1]{Shuanghou Deng}[type=author]
\credit{Funding acquisition, Writing - Review \& Editing, Supervision}

\author[inst1]{Jichang Chen}[type=author]
\credit{Data Curation, Writing - Review \& Editing}

\cortext[1]{Corresponding author}
\cortext[2]{Corresponding author}



\begin{abstract}
The present study is concerned with the peak acceleration $a_{z\mathrm{max}}$ occurring during the water impact of a symmetric wedge. This aspect can be important for design considerations of safe marine vehicles. The water-entry problem is \textcolor{black}{first studied} numerically using the finite-volume discretization of the incompressible Navier-Stokes equations and the volume-of-fluid method to capture the air-water interface. The choice of the mesh size and time-step is validated by comparison with experimental data of a free fall water-entry of a wedge. The key original contribution of the article concerns the derivation of a relationship for $a_{z\mathrm{max}}$ (as well as the correlated parameters $\upsilon_z^*$, $z^*$ and $t^*$ when $a_{z\mathrm{max}}$ occurs), the initial velocity $\upsilon_{z0}$, the deadrise angle $\beta$ and the mass $M$ of the wedge based on the transformation of von Karman's momentum theory which is extended with the inclusion of the pile-up effect. The pile-up coefficient $\gamma$, which has been proven dependent on $\beta$ in the case of water-entry with a constant velocity, is then investigated for the free fall motion and the dependence law derived from Dobrovol'skaya is still valid for $\beta$ varying from 10$^\circ$ to 45$^\circ$. Reasonable good theoretical estimates of $a_{z\mathrm{max}}$, $\upsilon_z^*$, $z^*$ and $t^*$ are provided for a relatively wide range of initial velocity, deadrise angle and mass using the extended von Karman's momentum theory which is the combination of the original von Karman's method and Dobrovol'skaya's solution and this theoretical approach can be extended to predict the kinematic parameters during the whole impacting phase.
\\

\noindent
© 2023. This manuscript version is made available under the CC-BY-NC-ND 4.0 license
https://creativecommons.org/licenses/by-nc-nd/4.0/
\end{abstract}


\begin{highlights}
\item The dependence law of the pile-up coefficient with respect to the deadrise angle is extended to the \textcolor{black}{free fall water-entry} of the wedge.
\item The initial velocity $\upsilon_{z0}$, the deadrise angle $\beta$ and the mass $M$ are the three key parameters for the wedge \textcolor{black}{free fall} water-entry condition.
\item A theoretical estimated approach is established and it can predict the kinematic parameters during impacting phase.
\item Considering the accuracy, a compromised value of the instantaneous Froude number, $Fr^* \approx 6.5$, can be used to identify the validity of the theoretical estimated approach.
\end{highlights}

\maketitle


\section{Introduction}
\label{sec:introduction}
Due to its complex fluid-structure interaction, water impacting problems have been widely concerned and investigated, not only in ocean engineering (such as ships and offshore structures), but also for the aircraft ditching/landing events, water-entry of projectiles and so on \citep{truscott2014water,chaudhry2020recent}. It is well-known that \cite{karman1929impact} first proposed an analytical estimation method based on a wedge-shaped water impact and introduced the method to settle the impact loads on seaplanes. Subsequently, a significant research work related to water impact has been carried out by theoretical, experimental, and numerical simulation approaches \citep{wagner1932phenomena,zhao1993water,scolan2001three,korobkin2004analytical,korobkin2006three,wu2014similarity,breton2020experimental,zekri2021gravity,delbuono2021water}. Numerous factors contribute \textcolor{black}{to} the impacting load including body shapes (symmetric/asymmetric), the relative motion between body and water (vertical/oblique/inclined entry), initial velocity profile (constant velocity/decelerated/accelerated motion), water-surface condition (calm water/wave) and many others. The effects determined by these factors on the kinematic parameters have been mostly discussed.

However, few investigations concern the general phenomenon of force history and related kinematic behaviours induced by the various initial factors during the impact, particularly on the most serious condition where the peak load occurs. Several studies have focused on the relationship between the maximum acceleration and initial parameters on free-falling water-entry. Among these studies, \textcolor{black}{\cite{gong2009water} carried out series of simulations on a wedge with various initial entering velocities by using a Smoothed Particle Hydrodynamics (SPH) model. They proposed fitting formulas for the maximum force acting on the wedge and the corresponding time when force reaches its peak with respect to the initial entering velocity. Moreover, for all cases with Froude number greater than 2, the maximum drag coefficients of the wedge turned out to be about 0.91.} In the work of \cite{abraham2014modeling}, the drag-coefficient of a sphere impacting the water surface was found to be independent of some investigated quantities, like the sphere velocity, surface tension, flow regime (laminar or turbulent) and Reynolds number. Hence, algebraic expressions of the drag coefficient versus the dimensionless depth have been established by two fitted polynomials. Effects of parametric variation, such as impact velocity, radius, and mass of the sphere on the impact force and the acceleration, have also been analyzed by \cite{yu2019parametric}. The peak value of the non-dimensional impact force has been found to be independent of the velocity and the radius, whereas it depends on the \textcolor{black}{sphere's mass}. In parallel, simplified expressions for the maximal force and acceleration have been \textcolor{black}{obtained by} fitting the relations between the peak value of the non-dimensional force and the non-dimensional mass. The relationships derived in \cite{yu2019parametric} have also been mentioned by other researchers’ work \citep{iafrati2019cavitation,iafrati2016experimental,wen2020impact,wang2021cfd,sheng2022acfd}.
\textcolor{black}{A similar study has been conducted experimentally by \cite{chen2022experiments} on the spherical ice ball impacting onto a rigid target, with the aim to establish expressions for the peak force and time of occurrence. Empirical equations are derived as functions of impact velocity, specimen radius, tensile strength of ice, ice density and elastic wave speed of ice by the dimensional analysis based on the Buckingham theorem}. Recently, the slamming force decomposition concept was numerically studied for 2D wedged shapes by \cite{sun2022slamming}. \textcolor{black}{The total slamming force can be expressed by three parts, such as speed, acceleration and gravity part, which can be evaluated by the corresponding coefficients depending on the penetration depth only. The improved force decomposition method predicts the slamming force accurately for a symmetric wedge (15, 20, 25 and 35$^\circ$ deadrise angle), a ship section and an asymmetric wedge (30$^\circ$ on the left and 20$^\circ$ on the right), but only available for various constant speed/acceleration combinations, not suitable for the free fall water-entry motion.}

It is worth noting that all the previous works have been trying to formulate functions of force, acceleration and other key parameters by the fitting approach and dimensionless method, and most of them only focus on the constant velocity water-entry. \textcolor{black}{In our previous study on free fall motion\citep{lu2022on},} it has been shown that the value of peak acceleration not only depends on the initial impact velocity, but also on the physical properties of the wedge, such as deadrise angle and mass, which has not been further investigated. The present study continues the previous one and it is mainly dedicated to numerical simulations of a two-dimensional symmetric wedge in free-falling water-entry in order to investigate and build up parametric equations, based on the extended von Karman's momentum theory. Compared with the previous work \citep{lu2022on}, the pile-up coefficient is further introduced into the theoretical formulations used to predict the maximal vertical acceleration and the corresponding vertical velocity, penetration depth and time when a wedge with a wide range geometric dataset impacts water. The adopted method can be extended to provide the force time history for the free-fall motion. The present work is organized as follows. Section \ref{sec:method} presents the methodology for the theoretical and numerical approaches, and describes the models and the computational setup; the main results are reported and discussed in \textcolor{black}{section} \ref{sec:results}; final conclusions are drawn in section \ref{sec:conclusion}.

\section{Methodology and Computational Setup}
\label{sec:method}
\subsection{Original von Karman's momentum theory and transformation}

\begin{figure}[hbt!]
\centering
\includegraphics[width=.35\textwidth]{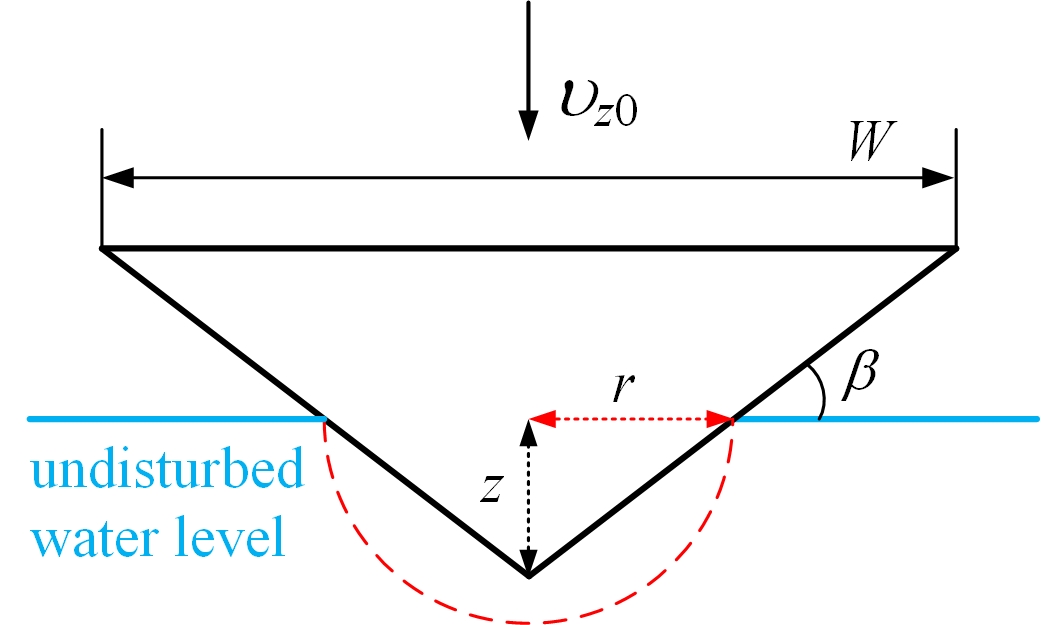}
\caption{Von Karman’s momentum approach.}
\label{fig:figVon}
\end{figure}

In our previous study \citep{lu2022on}, we proposed theoretical formulations, based on the original von Karman's momentum theory \citep{karman1929impact}, to predict the maximal vertical acceleration and the corresponding vertical velocity, penetration depth and time when the acceleration reaches its peak. In brief, by considering the momentum conservation at the beginning of the impact and the generic time $t$,
\begin{equation}
\label{eq:momentum}
M\upsilon_{z0}=(M+m_\mathrm{added}) \cdot \upsilon_{z}(t)
\end{equation}
where $M$ is the mass of the wedge per unit length (kg/m), $\upsilon_{z0}$ is the initial vertical impact velocity, $\upsilon_{z}(t)$ is the instantaneous velocity during the impact, and $m_\mathrm{added}=(\pi\rho r^2(t))/2$ is the added mass which is computed by using the flat-plate approximation (see Fig.~\ref{fig:figVon}), it is possible to retrieve the instantaneous velocity and acceleration of the impacting body as follows:

\begin{equation}
\label{eq:upsilon}
\upsilon_{z}(t)=\dfrac{2M\tan^2(\beta)\upsilon_{z0}}{2M\tan^2(\beta)+\pi\rho z^2(t)}
\end{equation}

\begin{equation}
\label{eq:a}
a_{z}(t)=\dfrac{\pi\rho z(t)}{M \upsilon_{z0} \tan^2(\beta)} \cdot \upsilon_{z}^3(t)
\end{equation}

Following \citep{lu2022on} the acceleration peak reads
\begin{equation}
\label{eq:amax}
a_{z\mathrm{max}}=\upsilon_{z0}^2 \left( \dfrac{5}{6} \right)^3 \dfrac{1}{\tan(\beta)} \sqrt{\dfrac{2\pi\rho}{5M}}
\end{equation}
and the corresponding vertical velocity, penetration depth and time taken when the acceleration attains its peak (and for this reason are here denoted with the superscript *) are
\begin{subequations}
\label{eq:threepara}
\begin{eqnarray}
\label{eq:threepara_z}
z^* &=& \sqrt{\dfrac{2M}{5\pi\rho}} \tan(\beta)\\
\label{eq:threepara_v}
\upsilon_z^* &=& \dfrac{5}{6} \upsilon_{z0} \\
\label{eq:threepara_t}
t^* &=& \dfrac{1}{\upsilon_{z0}}\dfrac{16}{15} \sqrt{\dfrac{2M}{5\pi\rho}} \tan(\beta) 
\end{eqnarray}
\end{subequations}
It is worth noting that we define the positive direction of acceleration upwards, while the vertical velocity and penetration depth are positive downwards. 

In Eqs.~\eqref{eq:amax} and \eqref{eq:threepara} there are several parameters which can affect the theoretical formulations. The significant role played by the initial vertical velocity, $\upsilon_{z0}$, has been deeply investigated and discussed in the previous study \citep{lu2022on}. It is worth investigating the effects of deadrise angle $\beta$ and mass $M$ on those relations.

\subsection{Extended von Karman's momentum theory}

\textcolor{black}{In the previous study, some differences observed between numerical results and theoretical formulations on the corresponding penetration depth $z^*$ (Fig. 11b in \cite{lu2022on})} have been ascribed to the pile-up effect which is neglected in the definition of the added mass based on original von Karman's theory. For this reason, a new formulation is here proposed \textcolor{black}{to take into account the pile-up effect.}

Considering the pile-up coefficient, $\gamma$, the added mass $m_\mathrm{added}$ can be rewritten by $m_\mathrm{added}=(\pi\rho\gamma^2 r^2(t))/2$. By using this new expression, the instantaneous velocity and acceleration of the impacting body become:

\begin{subequations}
\label{eq:upsilonGamma}
\begin{eqnarray}
\label{eq:upsilonGammaVt}
\upsilon_{z}(t) &=& \dfrac{2M\tan^2(\beta)\upsilon_{z0}}{2M\tan^2(\beta)+\pi\rho\gamma^2 z^2(t)}\\
\label{eq:upsilonGammaAt}
a_{z}(t) &=& \dfrac{\pi\rho\gamma \textcolor{black}{^2} z(t)}{M \upsilon_{z0} \tan^2(\beta)} \cdot \upsilon_{z}^3(t)
\end{eqnarray}
\end{subequations}
Similarly to what done before, it is possible to compute the peak of the impact acceleration and the corresponding penetration depth, velocity and time: 
\begin{subequations}
\begin{eqnarray}
\label{eq:threeparaGammaA}
a_{z\mathrm{max}} &=& \upsilon_{z0}^2 \left( \dfrac{5}{6} \right)^3 \dfrac{\gamma}{\tan(\beta)}\sqrt{\dfrac{2\pi\rho}{5M}}\\
\label{eq:threeparaGammaZ}
z^* &=& \sqrt{\dfrac{2M}{5\pi\rho\gamma^2}} \tan(\beta)\\
\label{eq:threeparaGammaV}
\upsilon_z^* &=& \dfrac{5}{6} \upsilon_{z0}\\
\label{eq:threeparaGammaT}
t^* &=& \dfrac{1}{\upsilon_{z0}}\dfrac{16}{15} \sqrt{\dfrac{2M}{5\pi\rho\gamma^2}} \tan(\beta)
\end{eqnarray}
\label{eq:threeparaGamma}
\end{subequations}

Moreover, a new expression regarding the pile-up coefficient $\gamma$ can be obtained from Eq.~\eqref{eq:threeparaGammaA}:

\begin{equation}
\label{eq:gamma}
a_{z\mathrm{max}}=\upsilon_{z0}^2 \left( \dfrac{5}{6} \right)^3 \dfrac{\gamma}{\tan(\beta)} \sqrt{\dfrac{2\pi\rho}{5M}}
\longrightarrow
\gamma=\dfrac{a_{z\mathrm{max}}}{\upsilon_{z0}^2 \left( \dfrac{5}{6} \right)^3 \sqrt{\dfrac{2\pi\rho}{5M}}} \cdot \tan(\beta)
\end{equation}

\subsection{Numerical method}

The theoretical formulation expressed by equations \eqref{eq:amax}-\eqref{eq:threepara} and \eqref{eq:threeparaGamma}-\eqref{eq:gamma} is herein numerically tested and validated by using the commercial package Star CCM+ as the two-phase flow solver. In the present study the unsteady incompressible Reynolds-averaged Navier-Stokes equations with a standard $k-\omega$ two-equation turbulence model are solved by the finite volume method. The governing equations for the continuity condition and the momentum conservation condition can be written as:

\begin{equation}
\label{eq:continuity}
\dfrac{\partial u_i}{\partial x_i} = 0
\end{equation}

\begin{equation}
\label{eq:momentumConser}
\rho \dfrac{\partial u_i}{\partial t} + \rho \dfrac{\partial}{\partial x_j} (u_i u_j) = 
-\dfrac{\partial p}{\partial x_i} + \dfrac{\partial}{\partial x_j} (\nu \dfrac{\partial u_i}{\partial x_j} - \rho \overline{u'_i u'_j}) + \rho g_i
\end{equation}
where $u_i$ and $u_j$ ($i,j$ = 1,2,3) are the time averaged value of velocity, $x_i$ and $x_j$ ($i,j$ = 1,2,3) are the spatial coordinate components, $\rho$ is the fluid density, $p$ is the fluid pressure, $\nu$ is the fluid kinematic viscosity, $- \rho \overline{u'_i u'_j}$ is the Reynolds stress, and $g_i$ is the gravitational acceleration in the $i$-th direction.

The Semi-Implicit Pressure Linked Equations (SIMPLE) algorithm is employed to achieve an implicit coupling between pressure and velocity, and the gradient is reconstructed with the Green-Gauss Node Based method. The modified High Resolution Interface Capturing (HRIC) scheme is adopted for volume fraction transport. The convection terms, as well as the diffusion terms, are turned into algebraic parameters using second-order upwind and second-order central methods, respectively. The unsteady terms are discretized in the time domain by applying a second-order implicit scheme.

Volume of fluid (VOF) scheme, originally proposed by \cite{hirt1981volume}, is used in the present computational scheme to capture the water-air interface by introducing a variable, $\alpha_\mathrm{w}$, called the volume fraction of the water in the computational cell, which varies between 0 (air) and 1 (water) and is defined as:
\begin{equation}
\label{eq:alphaw}
\alpha_\mathrm{w}=V_\mathrm{w}/V,
\end{equation}
where $V_\mathrm{w}$ is the volume of water in the cell and $V$ is the volume of the cell. The volume fraction of the air in a cell can be computed as:
\begin{equation}
\label{eq:alphawa}
\alpha_\mathrm{a}=1-\alpha_\mathrm{w}.
\end{equation}
The volume of fraction is governed by the following equation:
\begin{equation}
\label{eq:alphawaTransp}
\dfrac{\partial \alpha}{\partial t} + u_i \dfrac{\partial \alpha}{\partial x_i} = 0.
\end{equation}
The effective value $\varphi_\mathrm{m}$ of any physical properties, such as density, viscosity, etc., of the mixture of water and air in the transport equations is determined by:
\begin{equation}
\label{eq:varphi}
\varphi_\mathrm{m}=\varphi_\mathrm{w}\alpha_\mathrm{w}+\varphi_\mathrm{a}(1-\alpha_\mathrm{w}).
\end{equation}

To accurately capture the dynamic behavior and the load generated by the water-entry process, the motion of the body \textcolor{black}{caused by} the fluid forces and moments at the surface is determined via a six degree-of-freedom (6DOF) model. The 6DOF model solves the equations for the rotation and translation of the center of mass of the object. The equation for the translation is formulated in the global inertial coordinate system:
\begin{equation}
\label{eq:Ftranslation}
M \cdot \dfrac{\mathrm{d} \boldsymbol\upsilon}{\mathrm{d}t}=\boldsymbol{F},
\end{equation}
and the rotation of the object is solved in the body local coordinate system by:
\begin{equation}
\label{eq:Mrotation}
\boldsymbol{L} \dfrac{\mathrm{d} \boldsymbol\omega}{\mathrm{d}t}+ \boldsymbol\omega \times \boldsymbol{L} \boldsymbol\omega=\boldsymbol{M}.
\end{equation}

Subsequently, a dynamic mesh strategy \citep{xiao2021hydrodynamic}, which moves the entire mesh rigidly along with the object at each time step according to the solution of the 6DOF model, is employed to deal with the relative motion between the fluid and the rigid body with on single grid domain. The detailed description about the mesh method can be found in our \textcolor{black}{previous} study \citep{lu2022on}.

\color{black}

\subsection{Numerical validation}
\begin{figure}[hbt!]
\centering
\includegraphics[width=.4\textwidth]{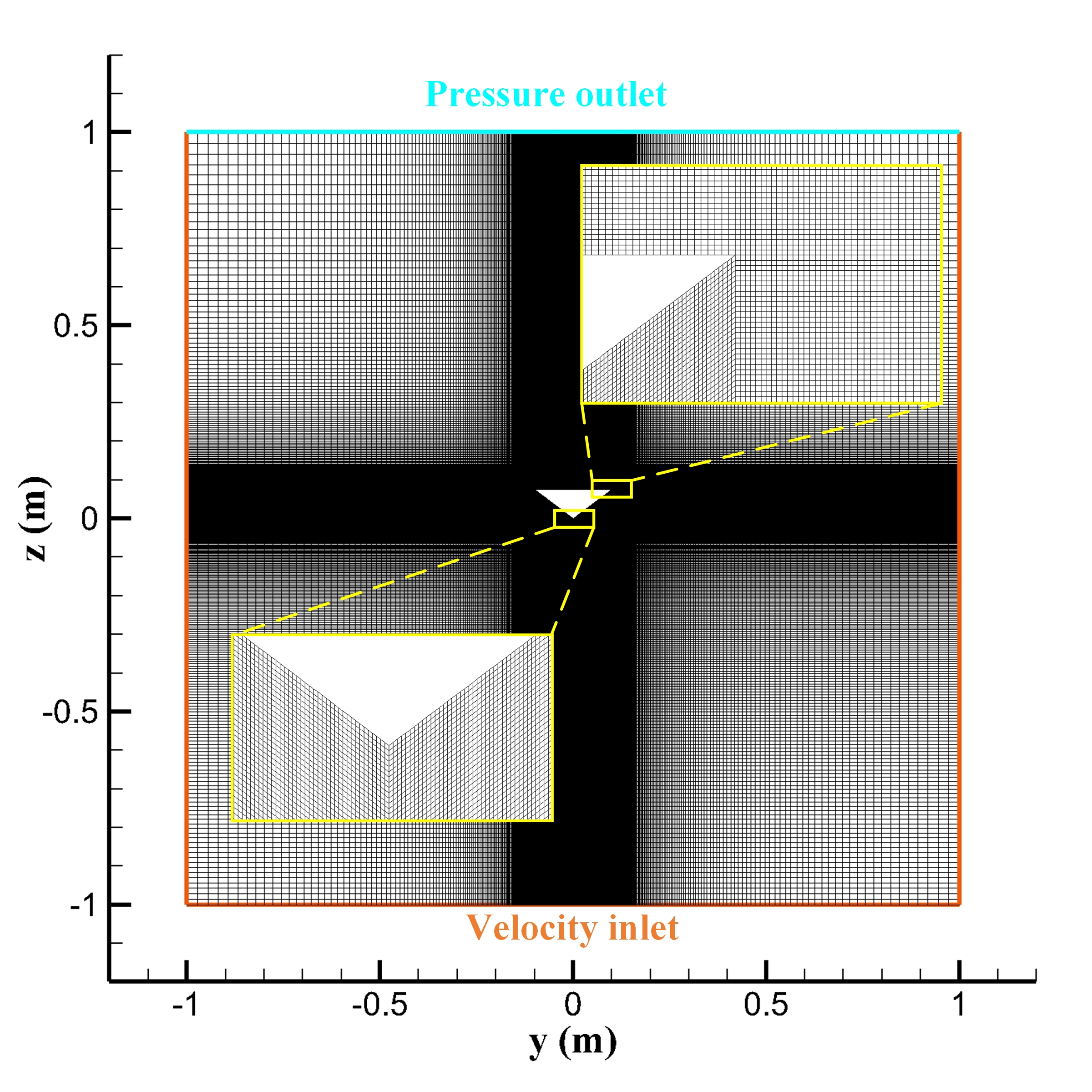}
\caption{Grid topology and density of wedge.}
\label{fig:mesh}
\end{figure}

First, the accuracy and efficiency of the numerical solver have been validated for the vertical water-entry of a symmetric wedge.
The simulated wedge has a width $W$ =0.2 m and a deadrise angle $\beta$ =37$^{\circ}$ referring to the experiment \textcolor{black}{of \cite{russo2018experimental}} and it is impacting with the symmetry axis oriented vertically, as seen in Fig.~\ref{fig:figVon}. Moreover, in order to carry out a relatively two-dimensional numerical simulation, the cell number in the y-direction (span-wise direction) is set to 1, and the cell size in the present study is 0.01 m. Fig.~\ref{fig:mesh} shows the mesh topology and grid density with two zoom-in views. The length of the square boundary is ten times the width of the wedge and the computational domain is discretized with structured quadrilateral grids. Inlet velocity conditions are enforced at the right, left and bottom boundaries, whereas pressure boundary condition is specified at the top boundary.

\begin{figure}[hbt!]
\centering
\begin{subfigure}{0.49\textwidth}
\includegraphics[width=\linewidth]{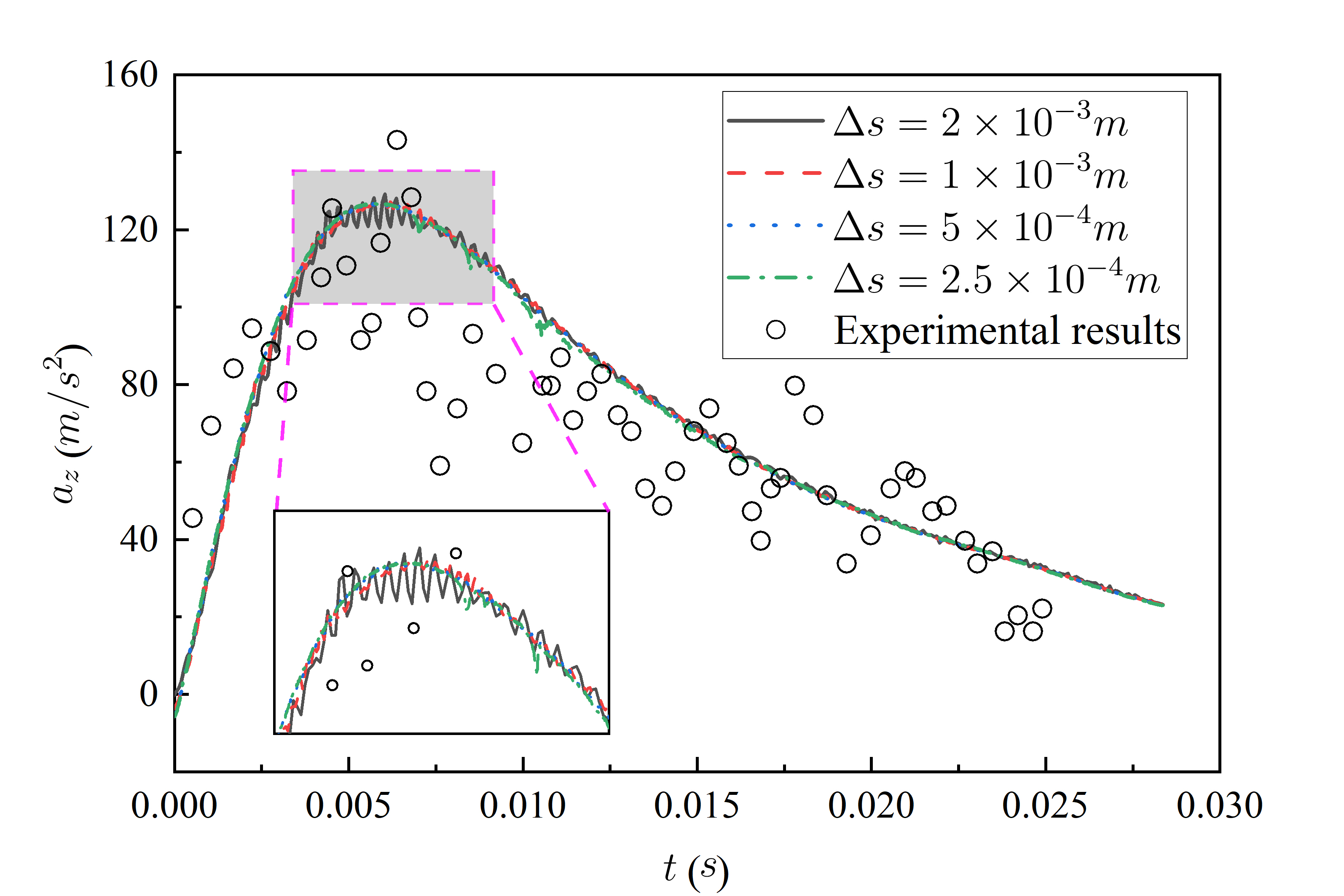} 
\caption{}
\label{fig:Grida}
\end{subfigure}
\begin{subfigure}{0.49\textwidth}
\includegraphics[width=\linewidth]{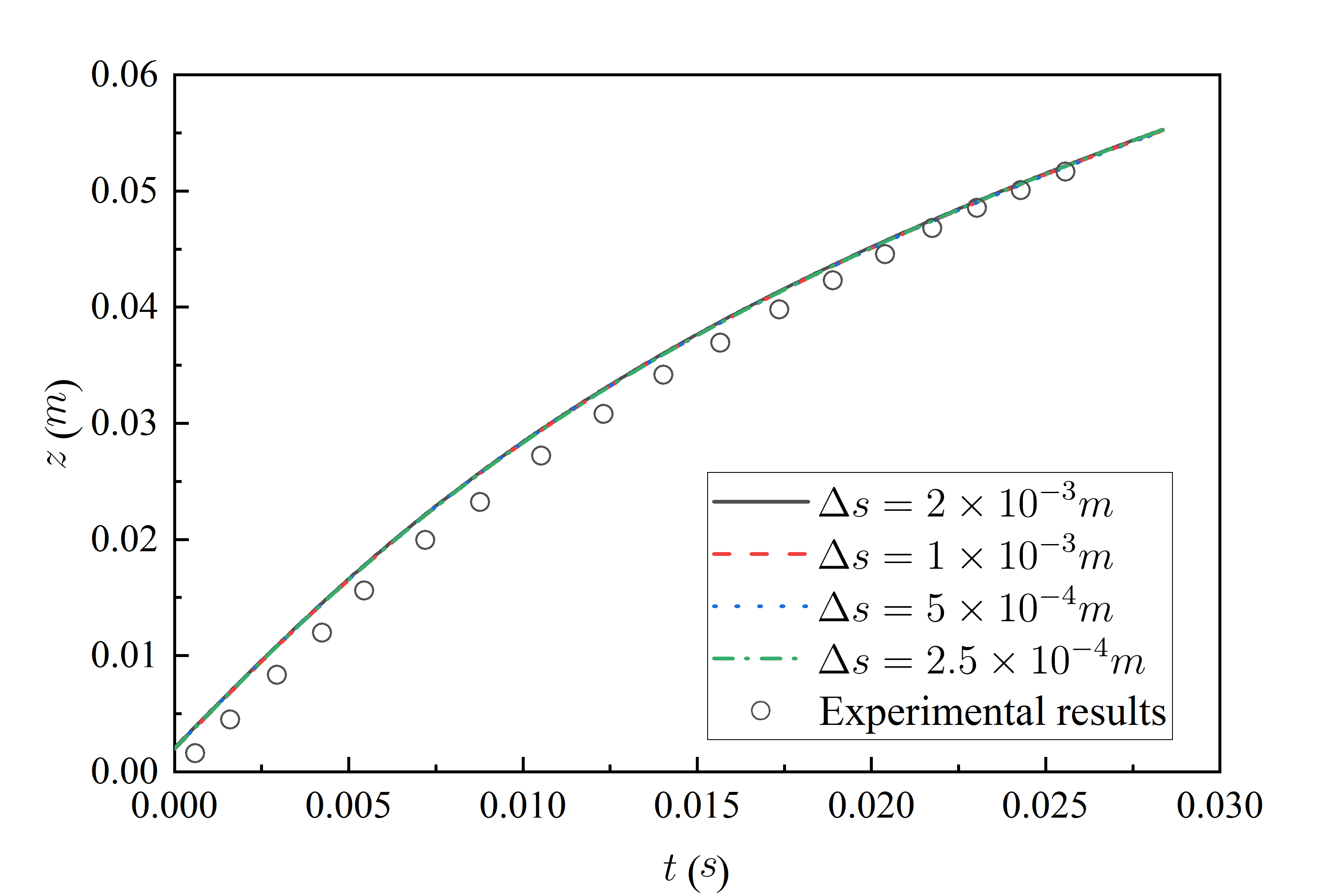}
\caption{}
\label{fig:Gridb}
\end{subfigure}
\caption{Grid independence tests with $\upsilon_{z0}=3.07$ m/s:(a) acceleration z; (b) displacement z.}
\label{fig:Grid}
\end{figure}

The aim of the validation is also investigating the convergence of the solution when increasing grid resolution and refining time step size. To assess the effects of spatial discretization, a mesh refinement study around the wedge has been conducted on four mesh systems with different resolution, as seen in Table~\ref{tab:Grid} in which $\Delta s$ represents the minimum size of mesh. Fig.~\ref{fig:Grid} shows the time histories of acceleration ($z$-direction) and displacement ($z$-direction) computed by the four meshes for the case of impacting with $\upsilon_{z0}=3.07$ m/s.
\textcolor{black}{Since the acceleration acting on the wedge is the most important parameter during free fall water-entry motion, particularly for the maximum acceleration closely related to the quantitative relationships (see Eq.~\eqref{eq:threeparaGamma}), numerical uncertainty due to discretization errors on the prediction of the maximum acceleration is considered and corresponding grid convergence index (GCI) method is derived \citep{celik2008procedure,islam2021assessment,wang2021cfd,han2022numerical}.} \textcolor{black}{Note that due to the unsatisfied oscillation on the grid 4 (very coarse), only three meshes (grid 1, 2 and 3, listed in Table~\ref{tab:Grid}) are used for the uncertainty study. The relevant results are presented in Table~\ref{tab:GridUncertainty}. We can now see that the solutions are in the asymptotic range of convergence, $GCI^{32}/(r^{p}GCI^{21}) \approx$1.000719, which is approximately one, and indicates that the solutions are well within the asymptotic range of convergence. Based on this study the maximum acceleration for the case $\upsilon_{z0}=3.07$ m/s is estimated to be 126.7123 $\mathrm{m/s^2}$ with an error band of 0.061$\%$. Furthermore, a good agreement with the experimental data \citep{russo2018experimental} is obtained and the results by the medium and fine mesh exhibit only minor differences. Thus the medium-size grid is chosen throughout the paper as a trade-off between computational expense and accuracy.}

\begin{table}[width=.6\linewidth, pos=hbt!]
\caption{Mesh resolution for grid convergence study with $\upsilon_{z0}=3.07$ m/s}
\centering
\label{tab:Grid}
\begin{tabular*}{\tblwidth}{@{}ccccc@{}}
\toprule
\textcolor{black}{No.}& Mesh& $\Delta s$, mm& Total cell number& Courant number\\
\midrule
\textcolor{black}{4}& Very coarse& 2& 183,300& 0.2\\
\textcolor{black}{3}& Coarse&      1& 231,750& 0.2\\
\textcolor{black}{2}& Medium&    0.5& 330,000& 0.2\\
\textcolor{black}{1}& Fine&     0.25& 525,000& 0.2\\
\bottomrule
\end{tabular*}
\end{table}

\begin{table}[width=.55\linewidth, pos=hbt!]
\caption{\textcolor{black}{Grid uncertainty estimation based on the maximum acceleration for the case $\upsilon_{z0}=3.07$ m/s}}
\centering
\label{tab:GridUncertainty}
\begin{tabular*}{\tblwidth}{@{}lcc@{}}
\toprule
Property& & $a_{z\mathrm{max}}, \mathrm{m/s^2}$\\
\midrule
Output values& $\phi_1$ (fine)& 126.6509\\
& $\phi_2$ (medium)& 126.5599\\
& $\phi_3$ (coarse)& 126.3344\\
Refinement ratio& $r_{21}$& 2\\
& $r_{32}$& 2\\
Order of accuracy& $p$& 1.31\\
Extrapolated value& $\phi_\mathrm{ext}^{21}$& 126.7123\\
& $\phi_\mathrm{ext}^{32}$& 126.7123\\
Approximate relative error& $e_\mathrm{a}^{21}$& 0.072$\%$\\
& $e_\mathrm{a}^{32}$& 0.178$\%$\\
Extrapolated relative error& $e_\mathrm{ext}^{21}$& 0.049$\%$\\
& $e_\mathrm{ext}^{32}$& 0.120$\%$\\
Grid convergence index (GCI)& $GCI^{21}$& 0.061$\%$\\
& $GCI^{32}$& 0.151$\%$\\
\bottomrule
\end{tabular*}
\end{table}

\begin{figure}[hbt!]
\centering
\begin{subfigure}{0.49\textwidth}
\includegraphics[width=\linewidth]{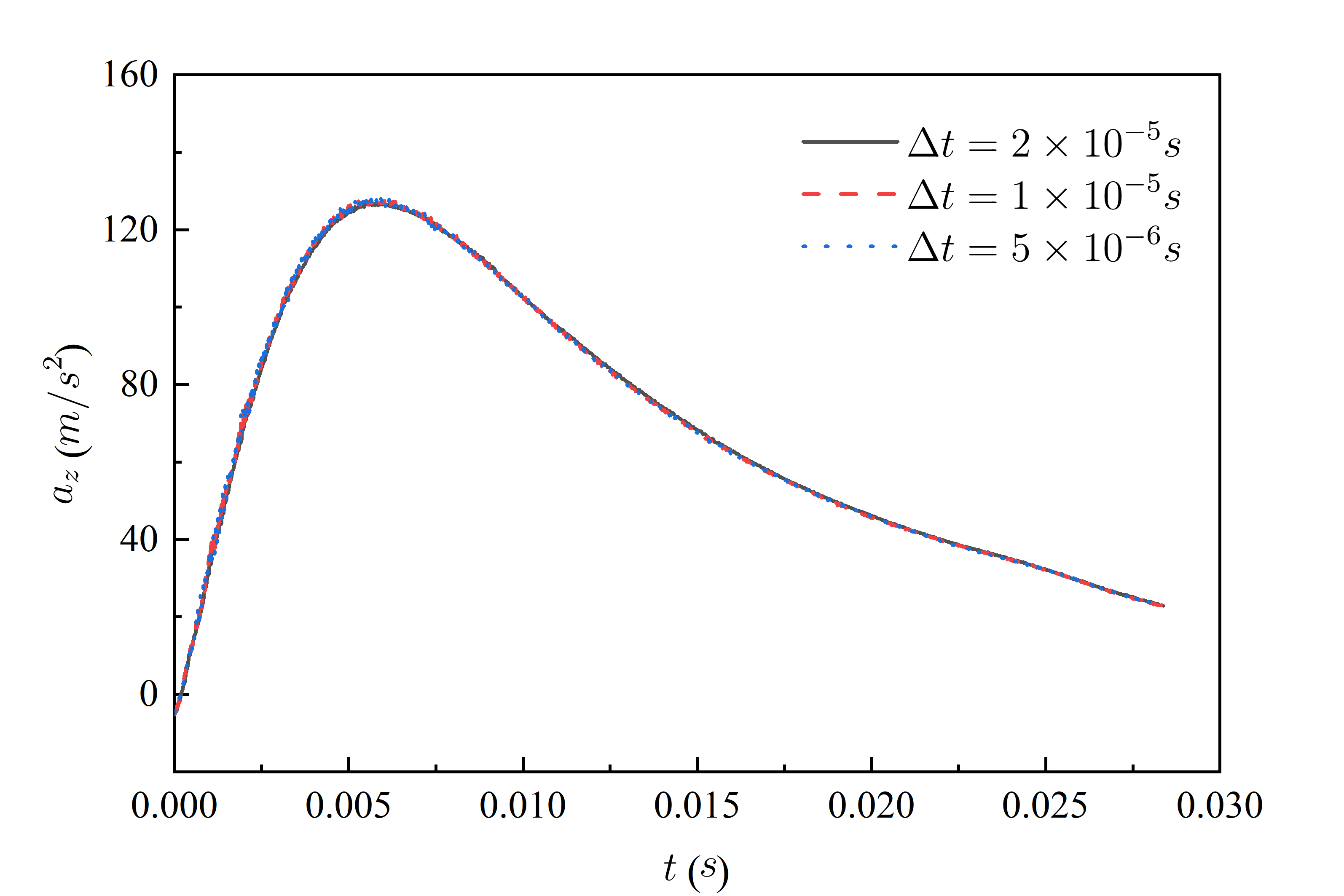} 
\caption{}
\label{fig:Timea}
\end{subfigure}
\begin{subfigure}{0.49\textwidth}
\includegraphics[width=\linewidth]{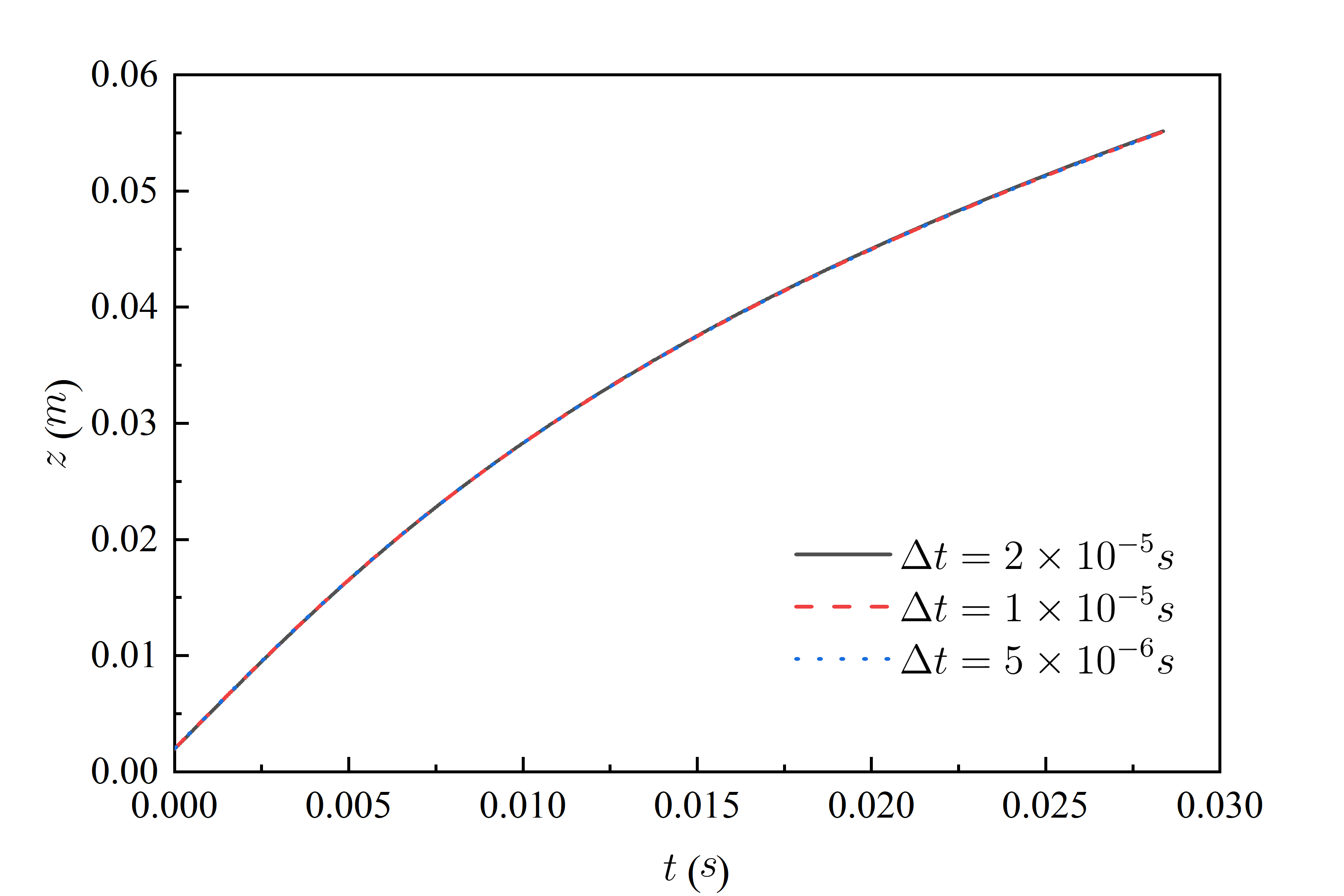}
\caption{}
\label{fig:Timeb}
\end{subfigure}
\caption{Time-step size refinement tests with $\upsilon_{z0}=3.07$ m/s:(a) acceleration z; (b) displacement z.}
\label{fig:Time}
\end{figure}

The time-step size refinement study is performed with three time-step sizes (see Table~\ref{tab:Time}) on the medium mesh for the same impacting condition. Fig.~\ref{fig:Time} shows the time-varying acceleration ($z$-direction) and displacement ($z$-direction). It can be seen that \textcolor{black}{nearly the same} variation trends of time-varying acceleration and displacement are captured by all the computations with these three time-step sizes. Considering both the efficiency and the accuracy of the simulation, $\Delta t = 1\times10^{-5}$s is selected as a reasonable alternative for wedge impacting cases.

\begin{table}[width=.35\linewidth, pos=hbt!]
\caption{Time-step size independence parameters with $\upsilon_{z0}=3.07$ m/s}
\centering
\label{tab:Time}
\begin{tabular*}{\tblwidth}{@{}ccc@{}}
\toprule
Mesh& $\Delta t, \times10^{-5}$s& Courant number\\
\midrule
Medium&    2&  0.2\\
Medium&    1&  0.1\\
Medium&  0.5& 0.05\\
\bottomrule
\end{tabular*}
\end{table}

\color{black}

\section{Results and Discussion}
\label{sec:results}
\begin{figure}[hbt!]
\centering
\includegraphics[width=.45\textwidth]{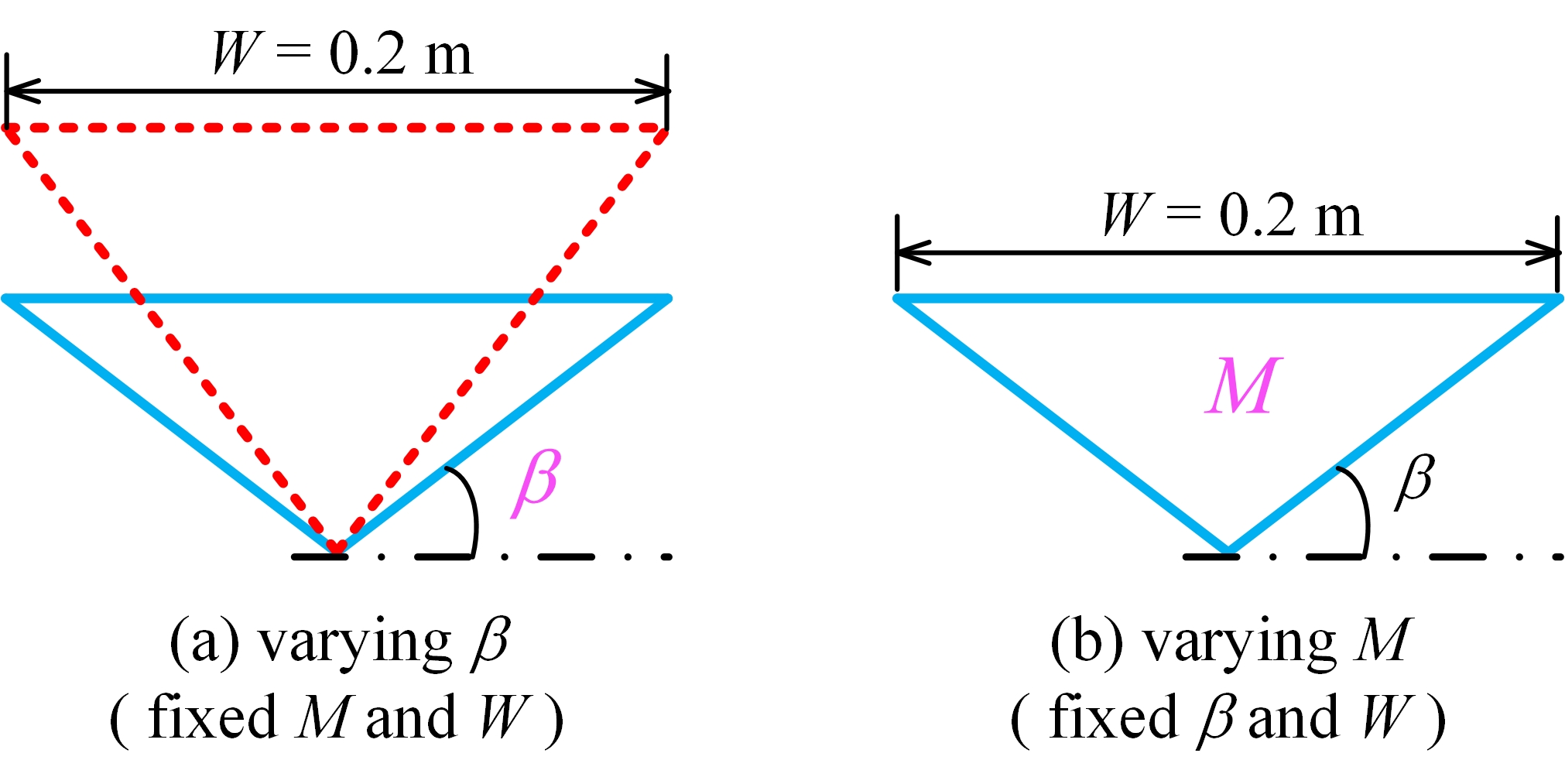}
\caption{Investigated parameters of the wedge.}
\label{fig:Config}
\end{figure}

In the present study, the simulation of 2D symmetric wedge vertical entry has been carried out to investigate the effect of the deadrise angle $\beta$ and mass $M$ on the dynamic behaviours of wedge, as shown in Fig.~\ref{fig:Config}.

\subsection{Effect of the deadrise angle}
Referring to the previous study \citep{lu2022on}, the theoretical estimates on the corresponding penetration depth $z^*$ and the ratio of velocity $\kappa$, \textcolor{black}{defined as $\kappa = \upsilon_{z}^{*}/\upsilon_{z0}$,} are nearly valid only when \textcolor{black}{$\upsilon_{z0}$ is greater than 2.95 m/s and 1.85 m/s respectively.} Thus, it is important to investigate the effect of deadrise angle on the theoretical equations (Eq.~\eqref{eq:amax} and \eqref{eq:threepara}) with a large and a small $\upsilon_{z0}$, chosen as 5.5 and 1 m/s, respectively. Note that the mass and the width of the wedge is kept constant for all the cases, that is \textcolor{black}{$M$ = 4.6842 kg/m} and $W$ = 0.2 m.

\subsubsection{High initial velocity}
\begin{figure}[hbt!]
\centering
\begin{subfigure}{0.49\textwidth}
\includegraphics[width=\linewidth]{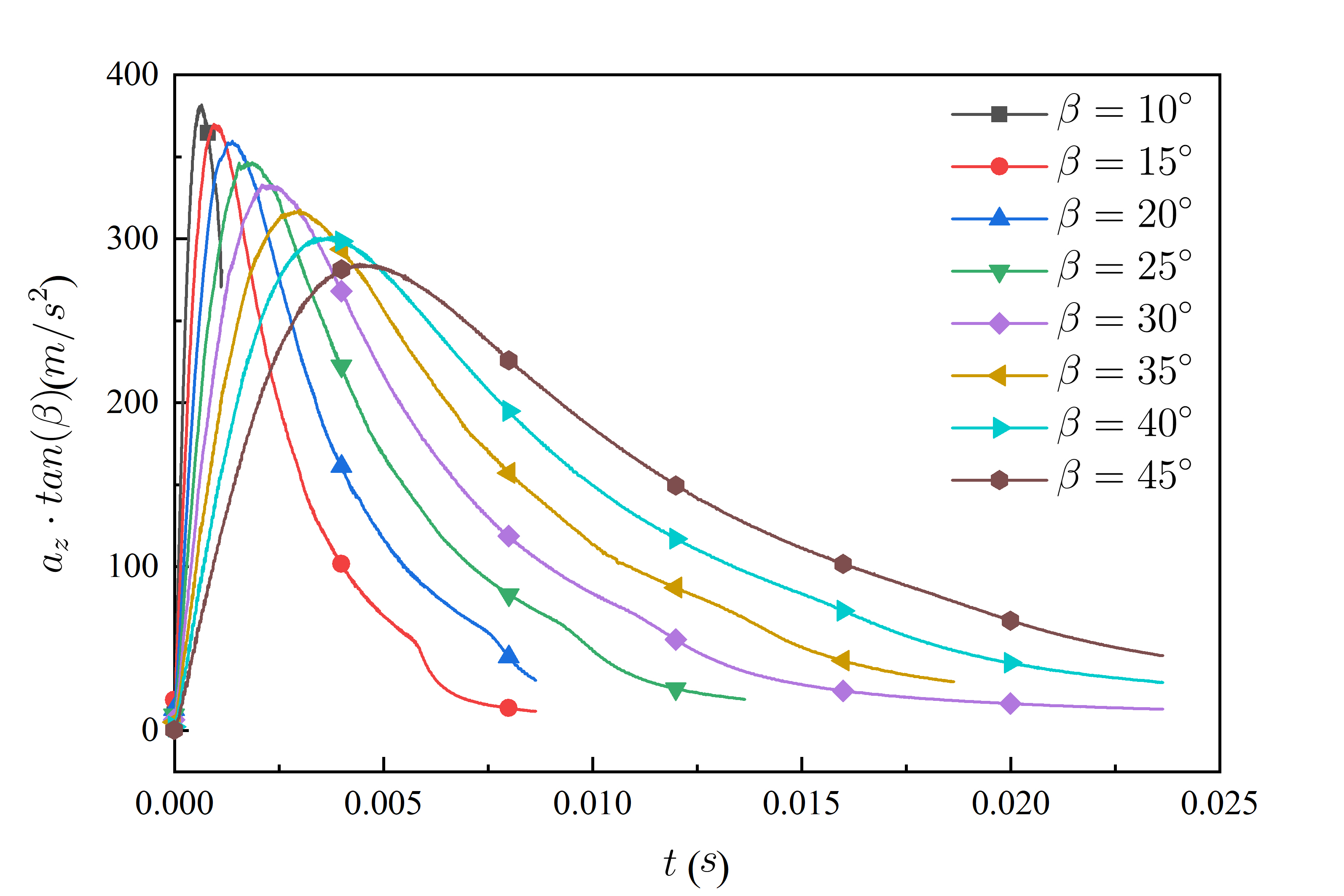} 
\caption{}
\label{fig:DifBeta5p5Aza}
\end{subfigure}
\begin{subfigure}{0.49\textwidth}
\includegraphics[width=\linewidth]{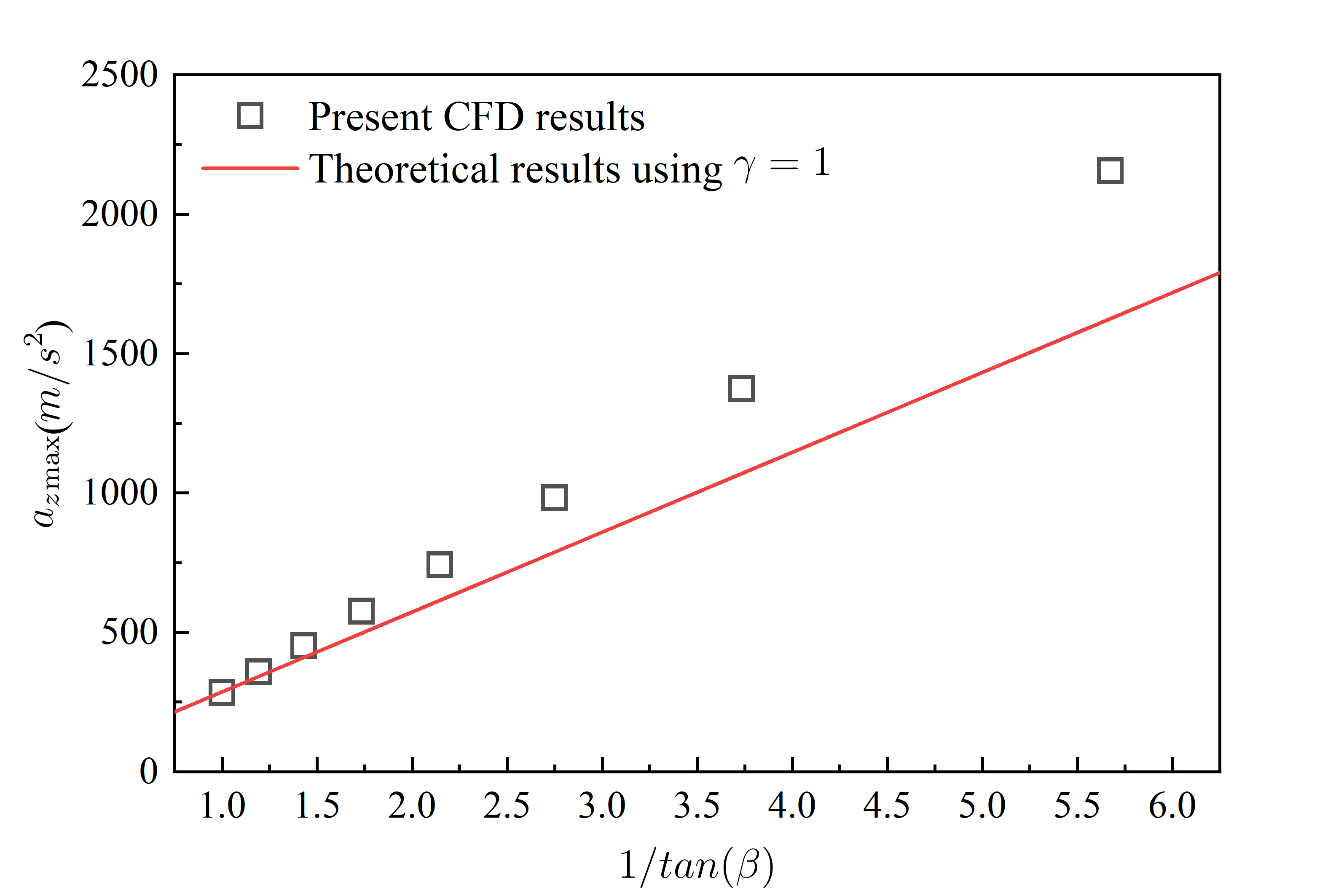}
\caption{}
\label{fig:DifBeta5p5Azb}
\end{subfigure}
\caption{Effect of deadrise angle $\beta$ in the case of $\upsilon_{z0}$=5.5 m/s: (a) variation of acceleration $a_z$ times $tan(\beta)$ versus time; (b) maximal acceleration $a_{z\mathrm{max}}$ versus $1/tan(\beta)$.}
\label{fig:DifBeta5p5Az}
\end{figure}

The case of initial velocity 5.5 m/s with different deadrise angle $\beta$, varying from $10^\circ$ to $45^\circ$, is first investigated. Considering Eq.~\eqref{eq:amax}, the term $a^* \tan(\beta)$ should be constant with varying deadrise angle $\beta$ for a given initial velocity and mass. Conversely, by looking at Fig.~\ref{fig:DifBeta5p5Aza}, an obvious difference can be observed among the peak positions of different deadrise angle conditions. \textcolor{black}{Moreover, by considering the evolution of the acceleration peak with respect to $1/\tan(\beta)$ in Fig.~\ref{fig:DifBeta5p5Azb}, a big deviation is observed with the results from the original von Karman's momentum theory (Eq.~\eqref{eq:amax}) with the decreasing of the deadrise angle. It should be noted that for a constant velocity water impact the decreasing of the deadrise angle strongly influences the pile-up effect and thus the value of the pile coefficient $\gamma$ \citep{mei1999on,iafrati2000hydroelastic,payne1994recent,dobrovol1969on}. The same seems to happen in the \textcolor{black}{free-falling} impacting scenario and it is then believed that the value of $\gamma$, which is $\gamma=1$ in the theory suggested by von Karman's, could explain the difference between the theoretical estimate and numerical result on the maximal acceleration value.} 

\begin{figure}[hbt!]
\centering
\includegraphics[width=.5\textwidth]{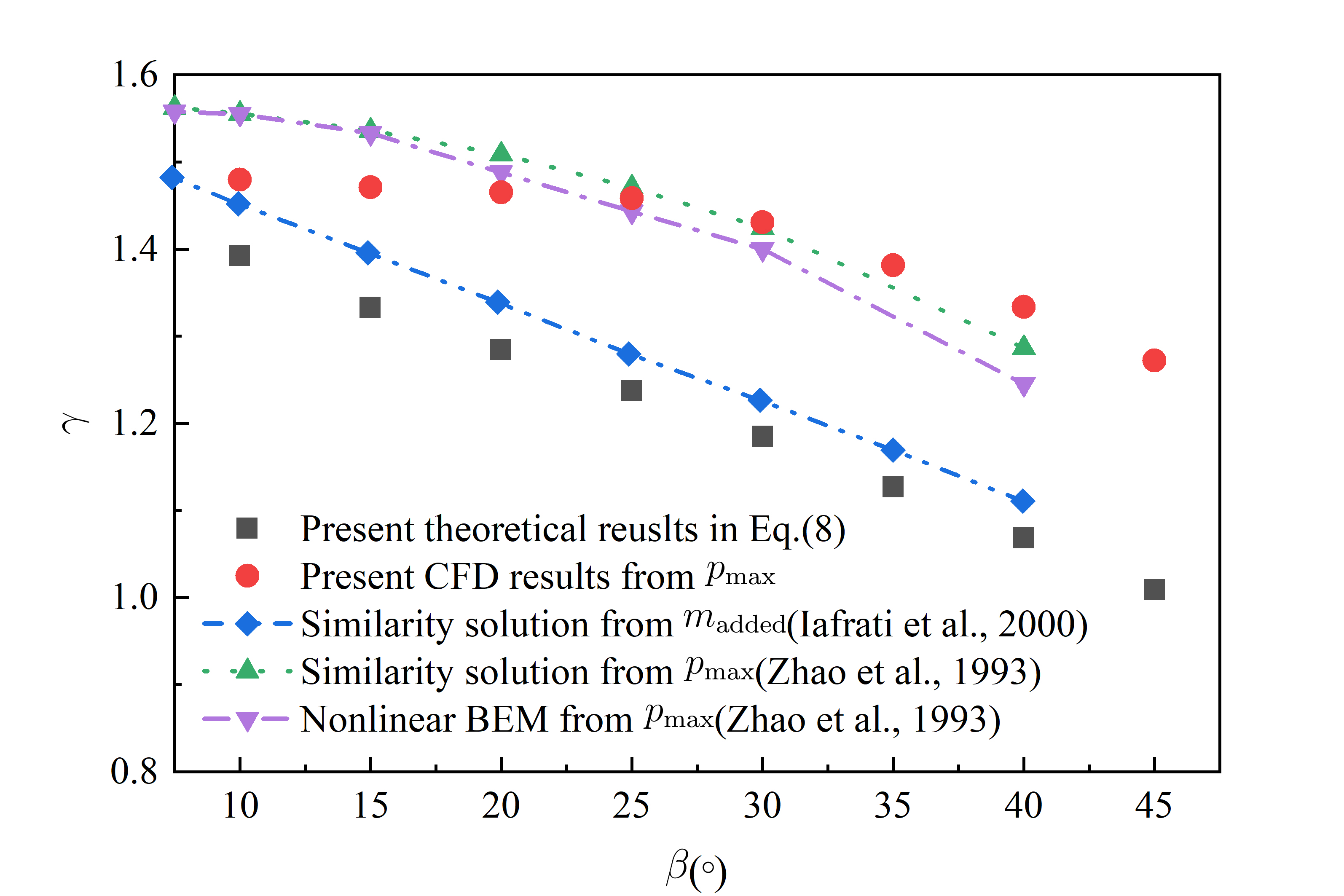}
\caption{Numerical estimate of $\gamma$ in the case of initial velocity $\upsilon_{z0}$ = 5.5 m/s compared with analytical and numerical solutions based on constant velocity water-entry.}
\label{fig:DifBeta5p5Gamma}
\end{figure}

\begin{figure}[hbt!]
\centering
\begin{subfigure}{0.49\textwidth}
\includegraphics[width=\linewidth]{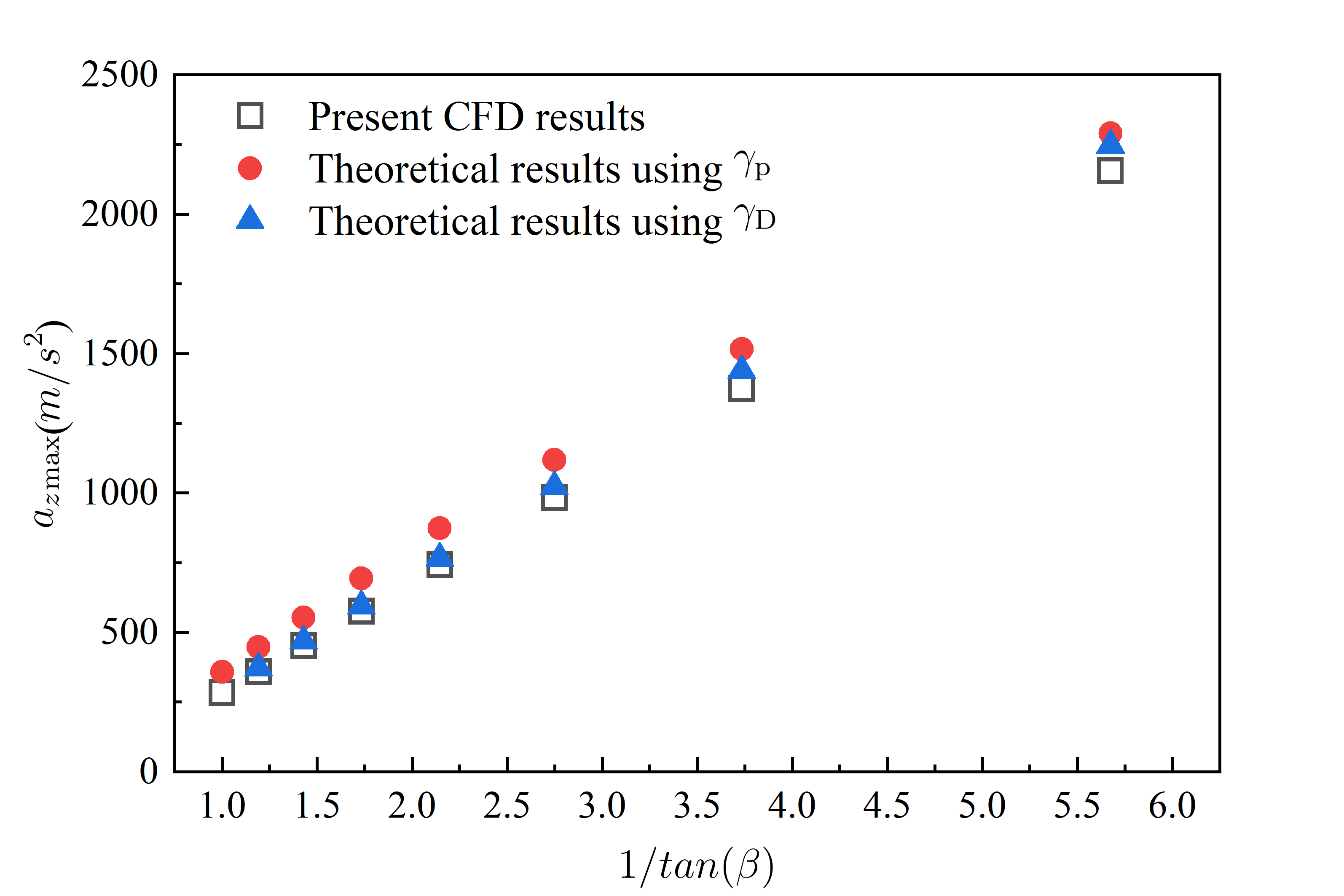} 
\caption{}
\label{fig:VerifyPDa}
\end{subfigure}
\begin{subfigure}{0.49\textwidth}
\includegraphics[width=\linewidth]{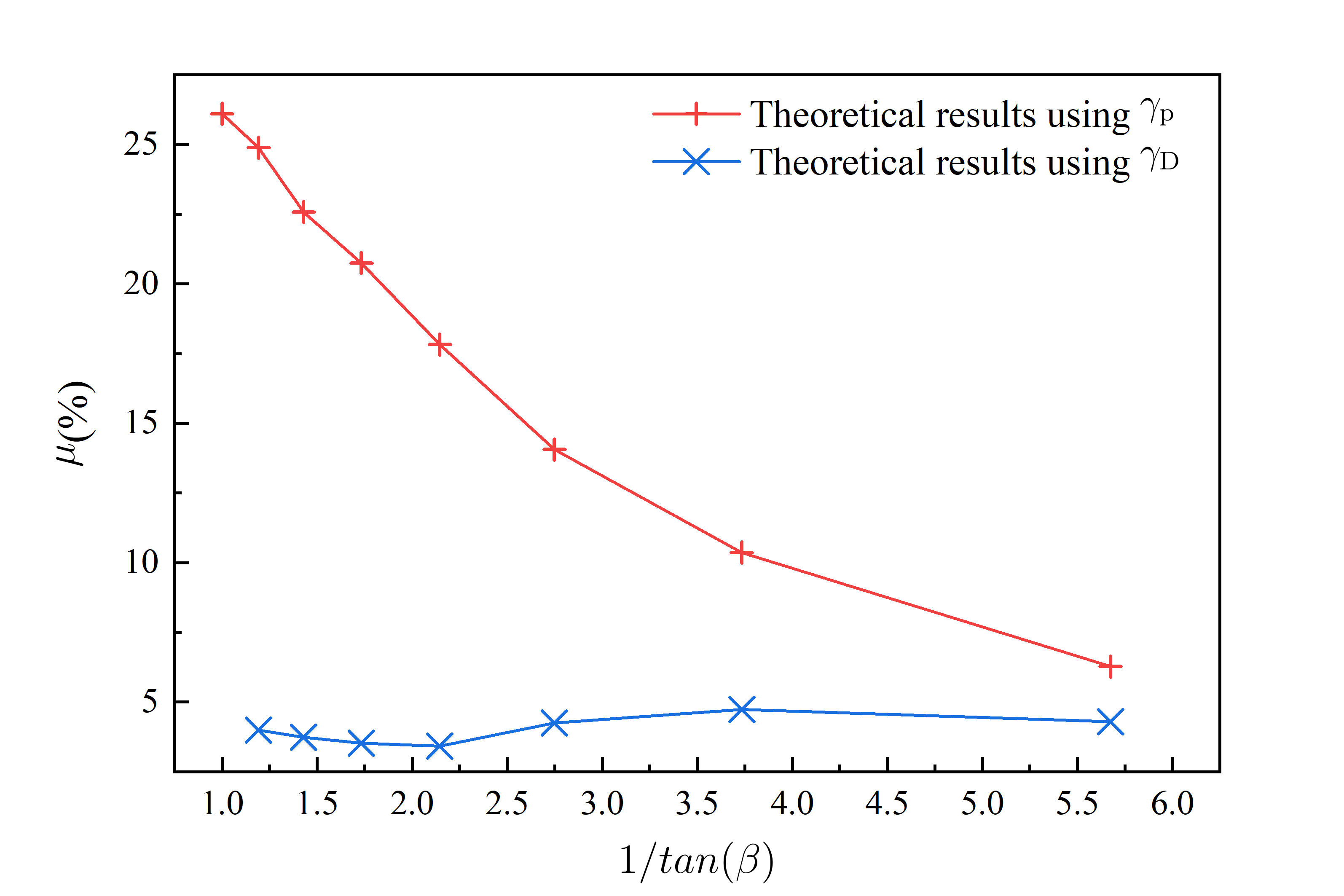}
\caption{}
\label{fig:VerifyPDb}
\end{subfigure}
\caption{Comparison of theoretical results using $\gamma_\mathrm{p}$ and $\gamma_\mathrm{D}$ on maximal acceleration $a_{z\mathrm{max}}$ in the case of $\upsilon_{z0}$=5.5 m/s: (a) variation of maximal acceleration $a_{z\mathrm{max}}$ versus $1/tan(\beta)$; (b) deviation $\mu$.}
\label{fig:VerifyPD}
\end{figure}

Using the numerical results of maximal acceleration shown in Fig.~\ref{fig:DifBeta5p5Azb}, the function of $\gamma$ and $\beta$ is then computed by using Eq.~\eqref{eq:gamma} and the results are displayed in Fig.~\ref{fig:DifBeta5p5Gamma}, compared with other solutions \citep{zhao1993water,iafrati2000hydroelastic,dobrovol1969on} obtained for impacting wedges with constant velocity. It is interesting to note that a similar behavior to the constant velocity case is observed in the free fall case. Indeed, the results, derived from the combination of von Karman's solution and numerical maximal acceleration value (see Eq.~\eqref{eq:gamma}), also exhibit a decay of $\gamma$ for increasing values of $\beta$ and tend to approach the similarity solution obtained by \cite{zhao1993water} by using the original formulation of \cite{dobrovol1969on}. The value of the pile-up coefficient provided by the similarity solution comes from the square root of non-dimensional added mass coefficient \citep{zhao1993water,iafrati2000hydroelastic}. Besides, the pile-up coefficient $\gamma$ can also be defined as $\gamma=r_\mathrm{peak}/r_\mathrm{still}$, where $r_\mathrm{peak}$ means the horizontal distance between the centerline of the wedge and the peak pressure position and $r_\mathrm{still}$ is the horizontal reference length measured from the wedge centerline to the intersection of the calm water surface and the wedge \textcolor{black}{\citep{chen2019wedge,kapsenberg2018phd}}. In Fig.~\ref{fig:DifBeta5p5Gamma}, a large difference can be observed when using the position of maximum pressure to compute the pile-up coefficient with the similarity solution, nonlinear boundary element method (BEM) \citep{zhao1993water} and the present CFD results. However, a decreasing trend of $\gamma$ is still valid increasing $\beta$ and when $\beta$ is greater than $20^\circ$, the values derived by the three different approaches are very close. It can be concluded that the pile-up coefficient $\gamma$, regardless of how it is evaluated, is dependent on the deadrise angle $\beta$, not only in the case of a constant velocity water impacting, but also for the free-falling motion, which has not been documented yet.
Then, using the two different expressions of $\gamma$ derived from Dobrovol'skaya's solution ($\gamma_\mathrm{D}$) and peak pressure position ($\gamma_\mathrm{p}$) into the Eq.~\eqref{eq:threeparaGammaA}, the values of maximal acceleration with varying $\beta$ are shown in Fig.~\ref{fig:VerifyPDa}. To evaluate the deviation on the maximal acceleration between the CFD results and theoretical results, a parameter $\mu$ is introduced, that can be expressed as: $\mu = \lvert ({C_\mathrm{theory} - C_\mathrm{CFD}})/{C_\mathrm{CFD}} \rvert \cdot 100\%$, where $C_\mathrm{theory}$ represents the results derived from the theoretical approach. As shown in Fig.~\ref{fig:VerifyPDb}, a good agreement can be achieved through $\gamma_\mathrm{D}$, with the deviation being below 5$\%$. Conversely, the agreement worsen when using $\gamma_\mathrm{p}$ and the deviation increases with the increasing of the deadrise angle.

%

\begin{figure}[hbt!]
\centering
\begin{subfigure}{0.49\textwidth}
\includegraphics[width=\linewidth]{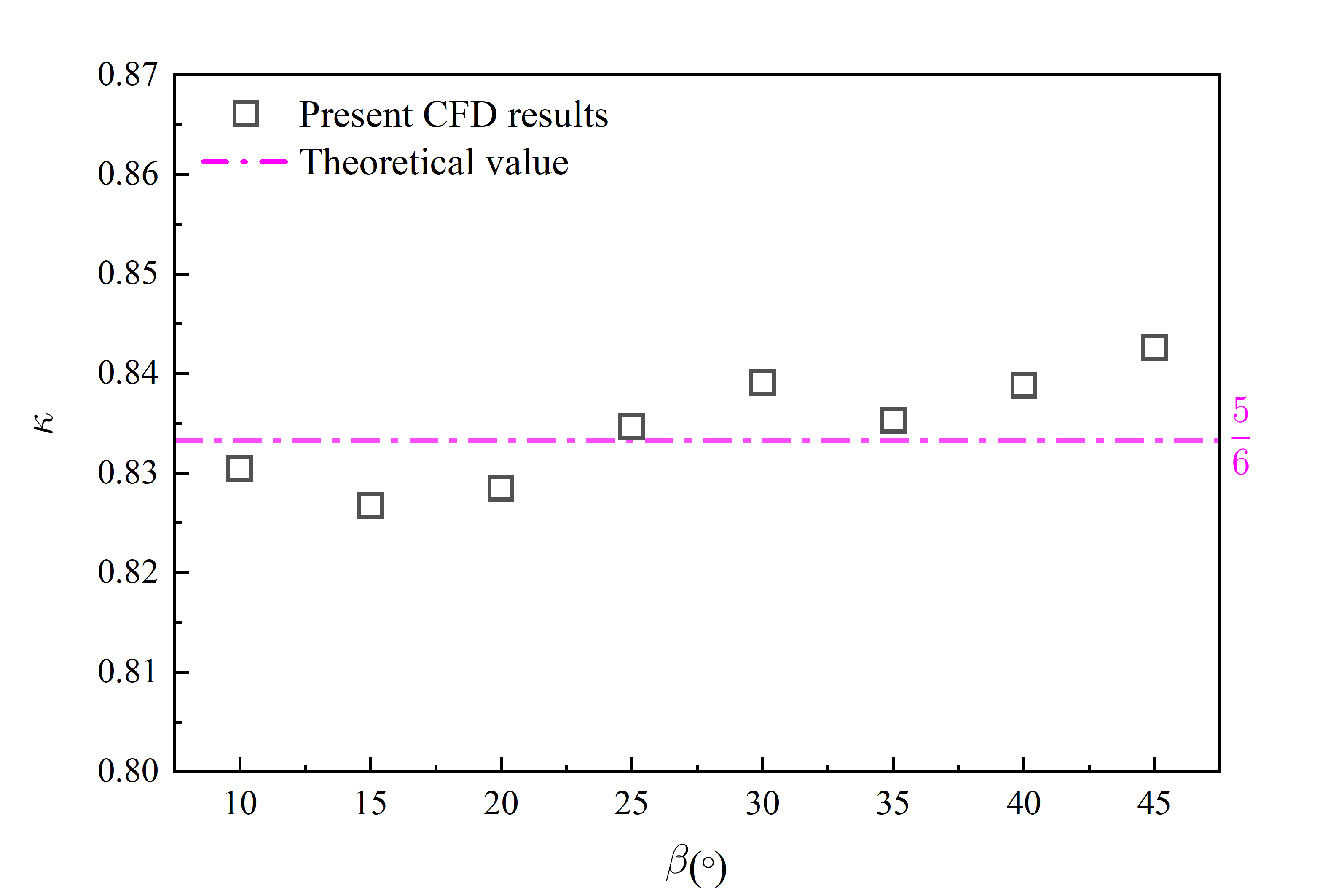}
\caption{}
\label{fig:DifBetaVz5p5TZVZKab}
\end{subfigure}
\begin{subfigure}{0.49\textwidth}
\includegraphics[width=\linewidth]{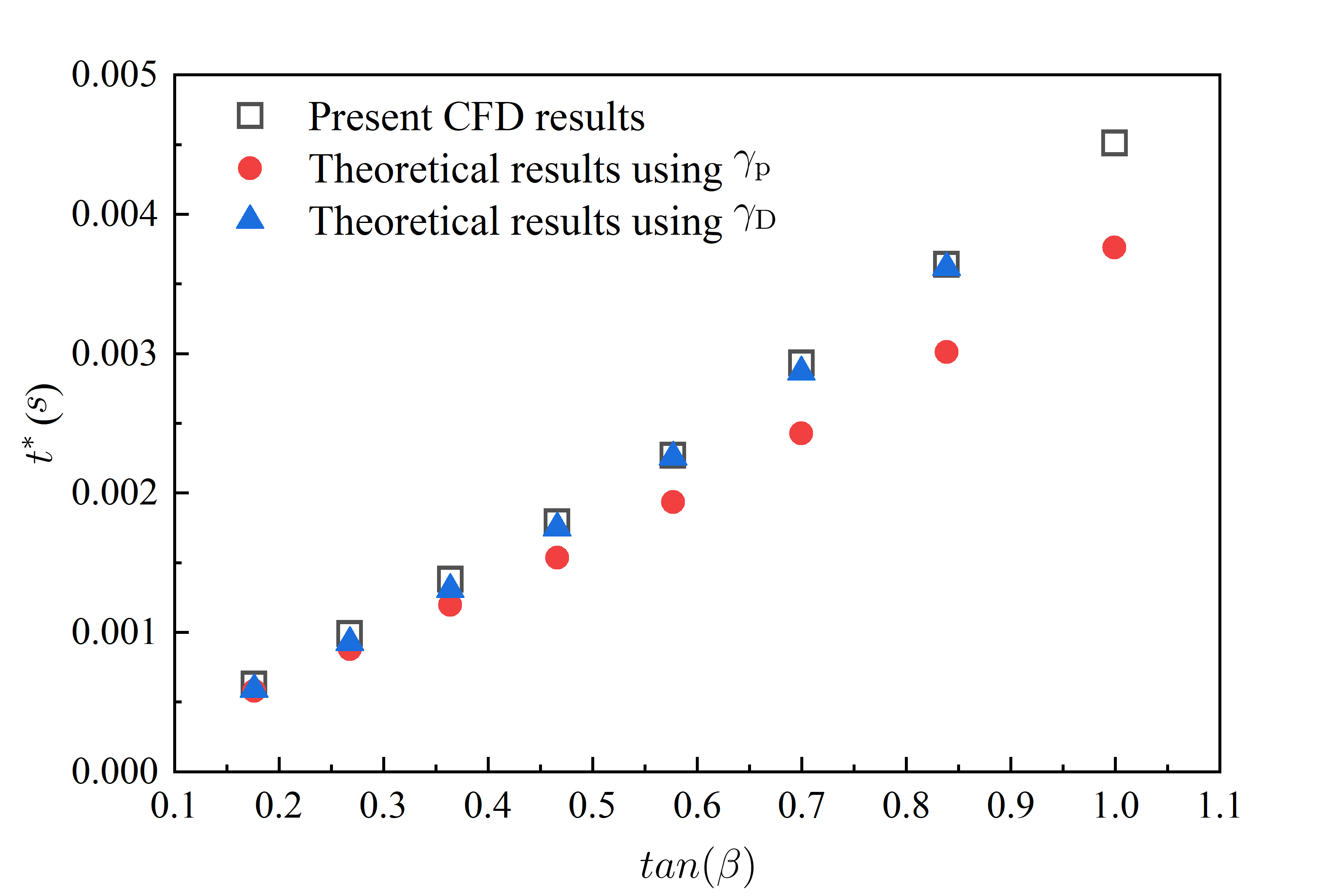} 
\caption{}
\label{fig:DifBetaVz5p5TZVZKac}
\end{subfigure}
\begin{subfigure}{0.49\textwidth}
\includegraphics[width=\linewidth]{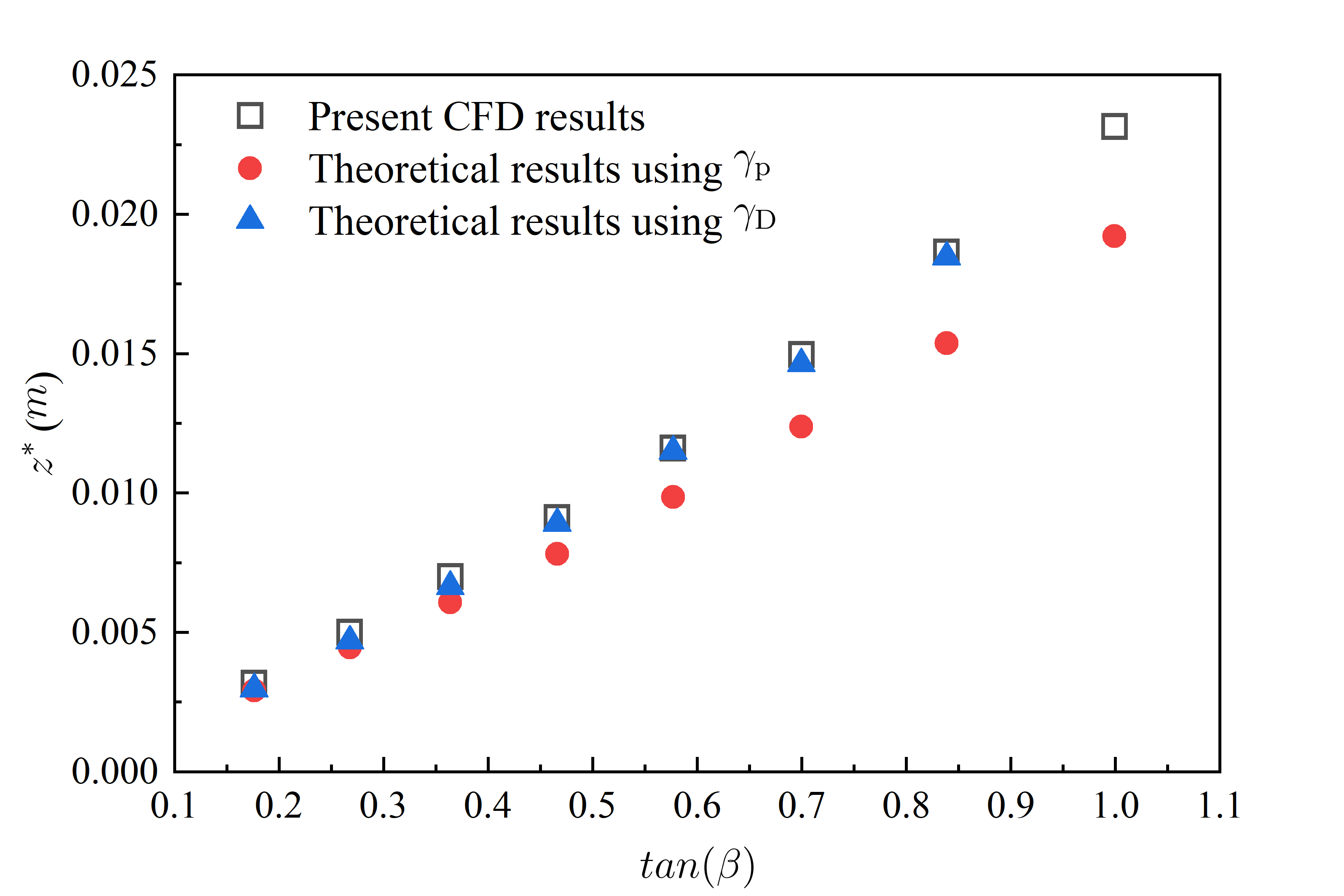}
\caption{}
\label{fig:DifBetaVz5p5TZVZKad}
\end{subfigure}
\caption{Effect of deadrise angle on variable dynamic parameters in the case of initial velocity $\upsilon_{z0}$ = 5.5 m/s: (a) $\kappa$; (b) $t^*$; (c) $z^*$.}
\label{fig:DifBetaVz5p5TZVZKa}
\end{figure}

In Fig.~\ref{fig:DifBetaVz5p5TZVZKa}, the three correlated variables are reported, viz., the ratio of velocity $\kappa$, time $t^*$ and penetration depth $z^*$ for the six cases. Note that $\kappa$ is defined as $\kappa=\upsilon_z^*/\upsilon_{z0}$. As it can be seen in Eq.~\eqref{eq:threeparaGammaV}, the value of $\upsilon_z^*$ is independent of the deadrise angle $\beta$ and only affected by the initial velocity, meaning that $\kappa$ is not affected by the deadrise angle $\beta$, that is similar to the solution in Fig.~\ref{fig:DifBetaVz5p5TZVZKab}. As it is shown in Fig.~\ref{fig:DifBetaVz5p5TZVZKab}, the results of $\kappa$ are very close to the theoretical estimated value with a root mean square error (RMSE) equal to 0.00536, indicating that $\upsilon_z^*$ is 5/6 times $\upsilon_{z0}$ in agreement with the theoretical estimate.

\begin{figure}[hbt!]
\centering
\includegraphics[width=.5\textwidth]{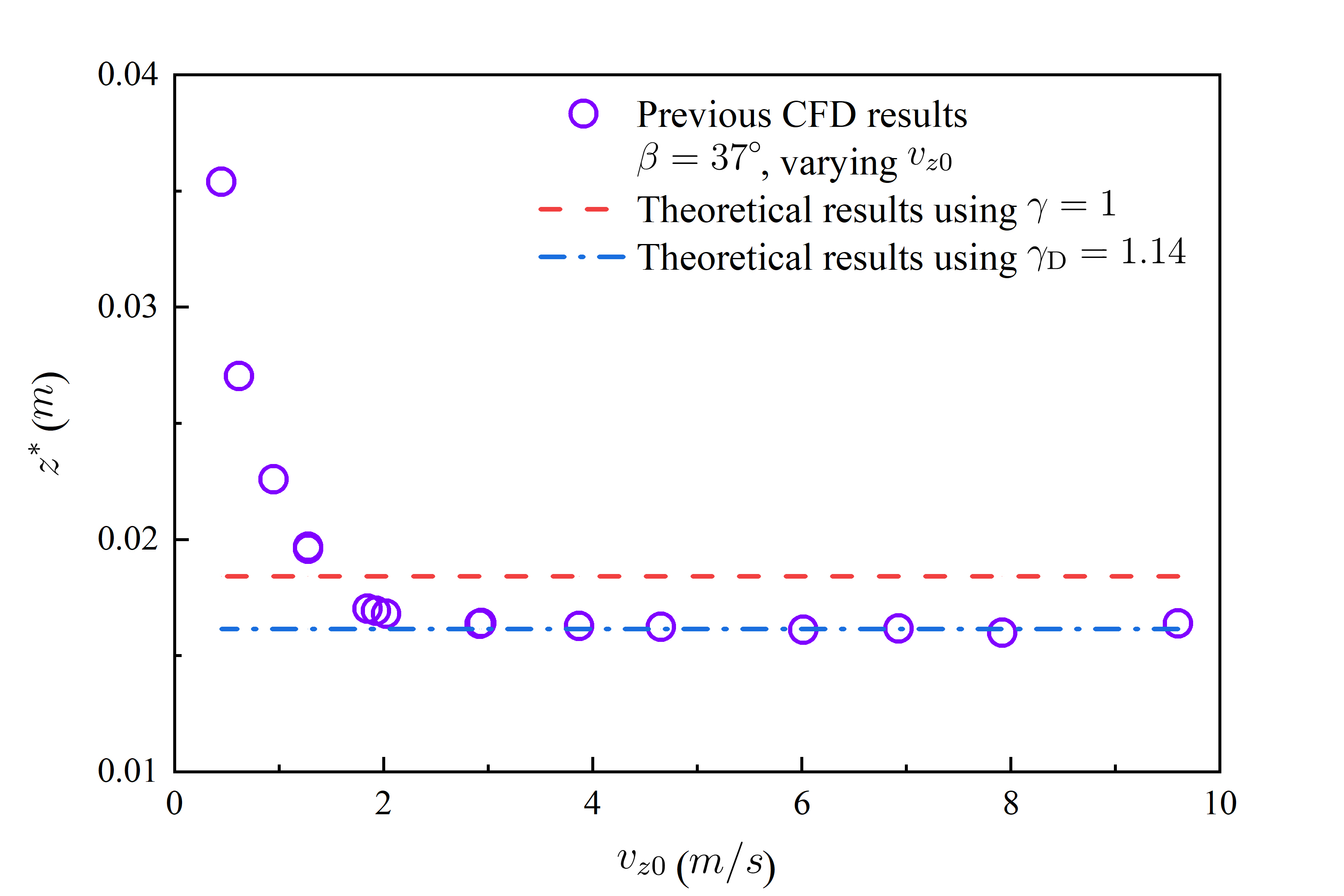}
\caption{Comparison of present theoretical results using $\gamma_\mathrm{D}$ and previous CFD results on $z^*$ versus $\upsilon_{z0}$ in the case of $\beta = 37^\circ$ with varying impact velocity.}
\label{fig:Verify37}
\end{figure}

Moving to Fig.~\ref{fig:DifBetaVz5p5TZVZKac} and Fig.~\ref{fig:DifBetaVz5p5TZVZKad}, $t^*$ and $z^*$ as linear functions of $tan(\beta)$ can be observed. The theoretical estimates using $\gamma_\mathrm{D}$ agree well with the numerical values, whereas the use of $\gamma_\mathrm{p}$ leads to an underestimation of those values. So far, it can be concluded that the quantitative relationships (see Eq.~\eqref{eq:threeparaGamma}) between the deadrise angle $\beta$ and the corresponding parameters ($a_z^*$, $z^*$, $v_z^*$ and $t^*$) are reliable in the case of initial velocity $\upsilon_{z0}$=5.5 m/s which is a relatively large initial impacting velocity. Furthermore, in the previous study for the case of $\beta = 37^\circ$ with varying initial velocity \citep{lu2022on}, a slight difference has been observed between the numerical asymptotic trend and the theoretical prediction in terms of $z^*$ when initial velocity increases, as shown here in Fig.~\ref{fig:Verify37}.
Conversely, when considering the pile-up effect by taking $\gamma_\mathrm{D}$ into Eq.~\eqref{eq:threeparaGammaZ}, the new theoretical formulation (blue dash dotted line) compares well with CFD results when increasing $\upsilon_{z0}$. So, it is confirmed that the present theoretical approach has a good accuracy on predicting the maximal acceleration and the correlated parameters, such as $z^*$, $v_z^*$ and $t^*$ when using quantitative relationships (Eq.~\eqref{eq:threeparaGamma}) associated with the pile-up coefficient $\gamma_\mathrm{D}$.

\subsubsection{Low initial velocity}
In addition to the analysis of the effect of the deadrise angle on the maximum acceleration $a_{z\mathrm{max}}$ and corresponding parameters $\upsilon_z^*$, $z^*$ and $t^*$ with a large initial velocity 5.5 m/s, \textcolor{black}{it is also worth investigating the case with a smaller initial velocity, in which gravity effect could be much more relevant \citep{lu2022on}.} Herein, the smaller initial velocity is setup as 1 m/s and the deadrise angle is varied from $10^\circ$ to $45^\circ$.

\begin{figure}[hbt!]
\centering
\includegraphics[width=0.5\textwidth]{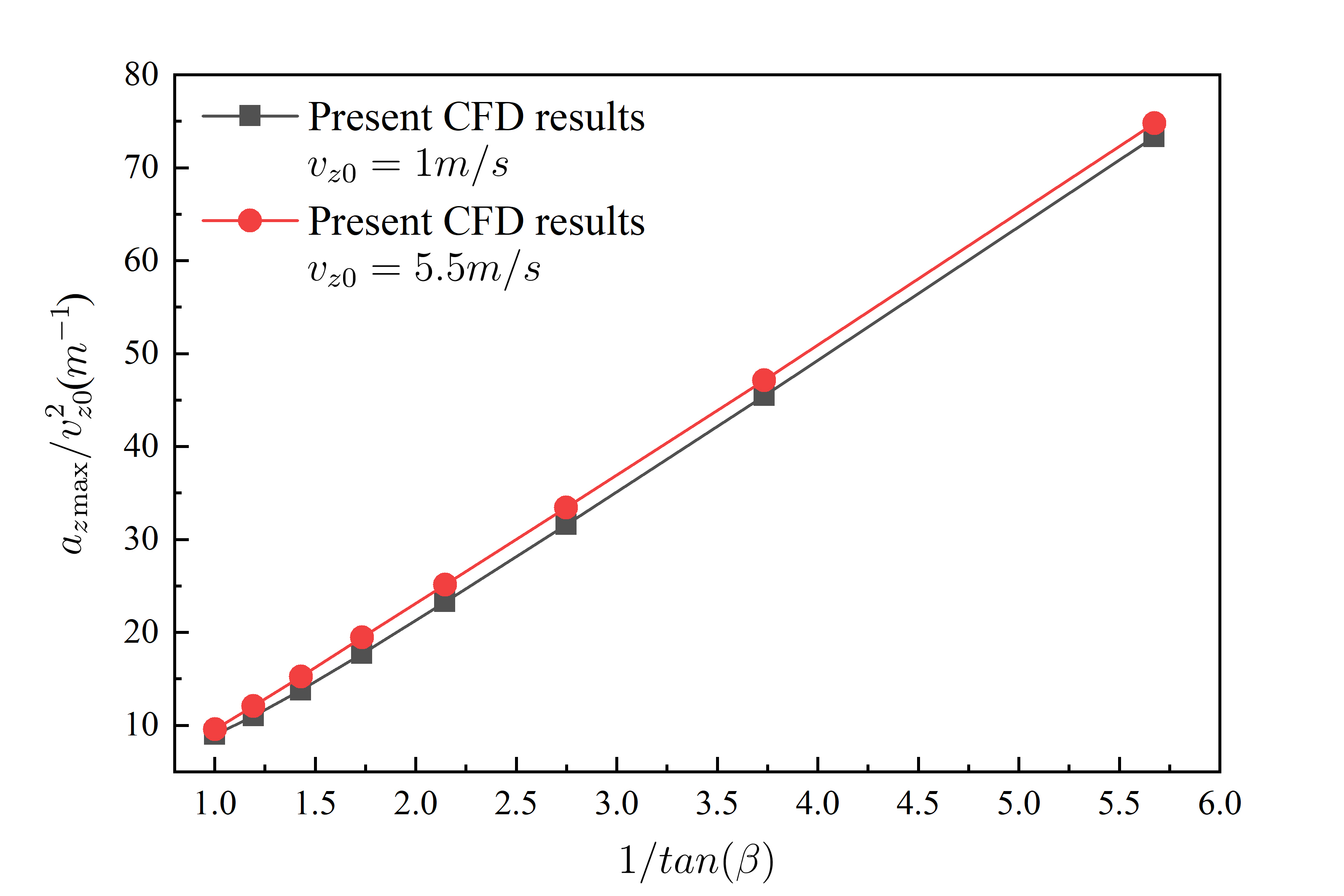} 
\caption{Effect of deadrise angle $\beta$ on the scaled maximal acceleration $a_{z\mathrm{max}}/\upsilon_{z0}^2$ versus $1/tan(\beta)$ in the case of $\upsilon_{z0}$=1 m/s compared with the case of 5.5 m/s.}
\label{fig:DifBeta1Az}
\end{figure}

\begin{figure}[hbt!]
\centering
\begin{subfigure}{0.49\textwidth}
\includegraphics[width=\linewidth]{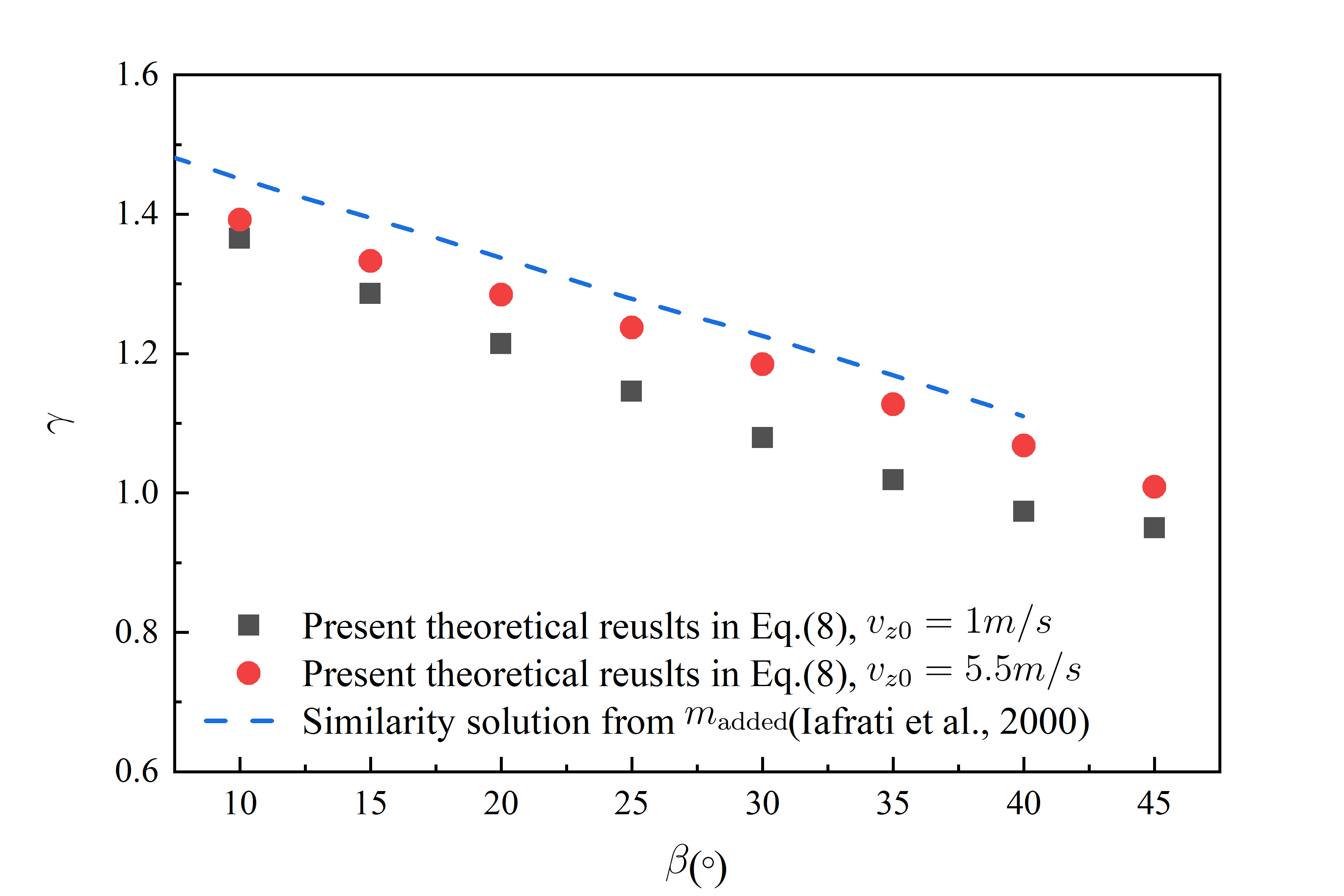} 
\caption{}
\label{fig:DifBeta1GammaCompaa}
\end{subfigure}
\begin{subfigure}{0.49\textwidth}
\includegraphics[width=\linewidth]{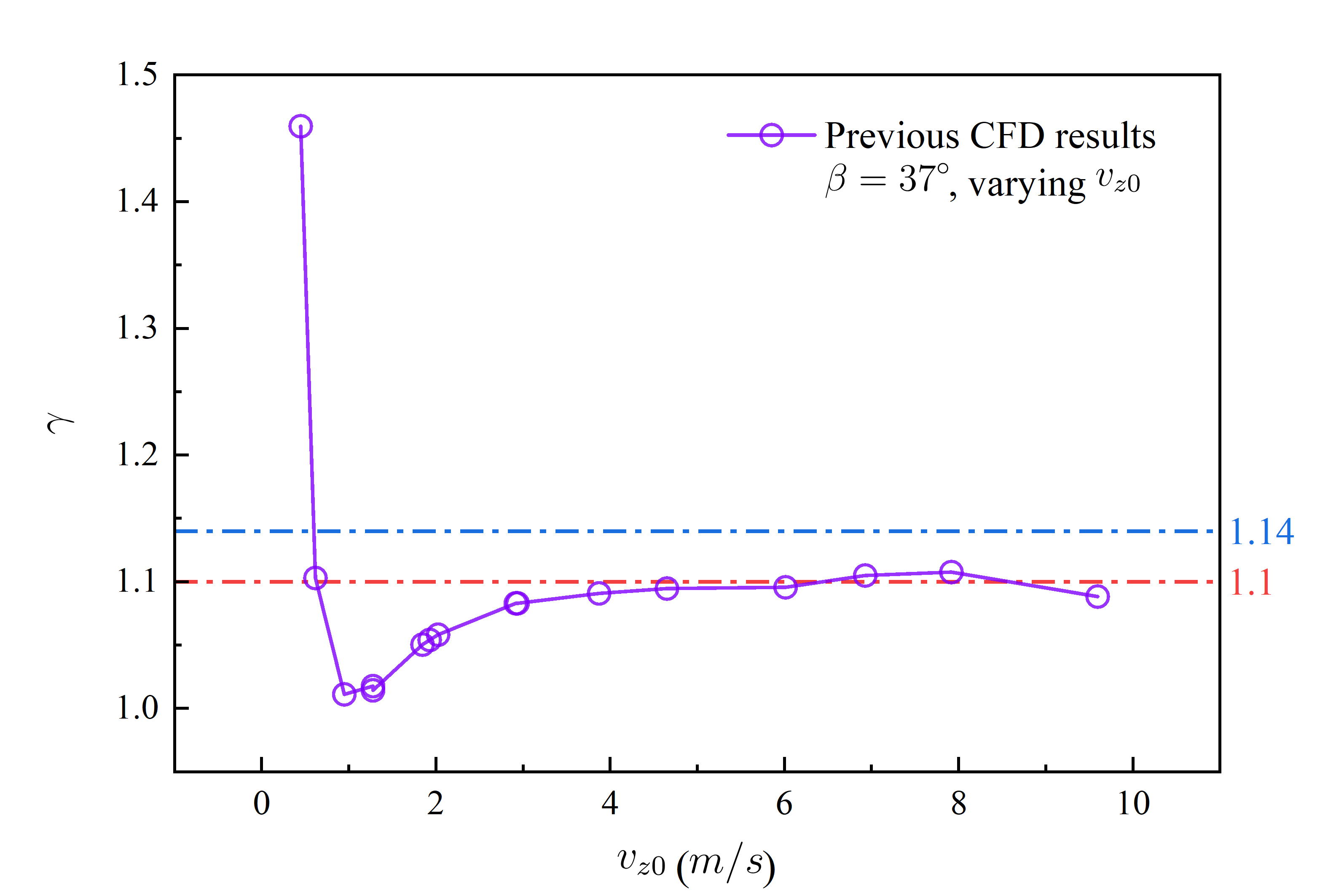}
\caption{}
\label{fig:DifBeta1GammaCompab}
\end{subfigure}
\caption{(a) Comparison of numerical estimate on $\gamma$ between the cases of $\upsilon_{z0}$=1 m/s and $\upsilon_{z0}$=5.5 m/s; (b) variation of $\gamma$ with the changing initial velocity for the case of $\beta=37^{\circ}$ \textcolor{black}{(bule dashed line: value extracted from the results of the similarity solution; red dashed line: value extracted from the case of $\upsilon_{z0}$=5.5 m/s)}.}
\label{fig:DifBeta1GammaCompa}
\end{figure}

Fig.~\ref{fig:DifBeta1Az} shows the comparison of maximal acceleration value, scaled with $\upsilon_{z0}^2$, with varying $\beta$ in the cases of $\upsilon_{z0}$=1 m/s and $\upsilon_{z0}$=5.5 m/s.
From Eq.~\eqref{eq:threeparaGammaA}, the two curves should be overlapped, \textcolor{black}{whereas} a little difference is observed.
Furthermore, as shown in Fig.~\ref{fig:DifBeta1GammaCompaa}, in the case of $\upsilon_{z0}$=1 m/s the data of the pile-up coefficient $\gamma$ (derived from Eq.~\eqref{eq:gamma} diminishes with the increase of deadrise angle $\beta$ and displays an overall reduction compared with the data from the case at $\upsilon_{z0}$=5.5 m/s. Additionally, the data of the previous study \citep{lu2022on} on the fixed deadrise angle $\beta$=37$^{\circ}$ with various initial velocity are shown in Fig.~\ref{fig:DifBeta1GammaCompab}, associated with two constant lines $\gamma_D$ = 1.14, extracted from the results of the similarity solution, and $\gamma$ = 1.1 which is the case of $\upsilon_{z0}$=5.5 m/s in Fig.~\ref{fig:DifBeta1GammaCompaa}. One can see a slight deviation of $\gamma$ when the initial velocity is larger than a certain value. Conversely, for the smaller initial velocity, the results of $\gamma$ exhibit firstly a decreasing trend, and then an increasing one (for $\upsilon_{z0} <$ 1 m/s) as the reduction of $a_{z\mathrm{max}}$ is much slower than the reduction of $\upsilon_{z0}$ causing the increasing of the ratio $a_{z\mathrm{max}}/\upsilon_{z0}^2$ in Eq.~\eqref{eq:gamma}, which provides the increase of $\gamma$. Hence, for the smaller initial velocity, where gravity effects play a very important role, the use of $\gamma_\mathrm{D}$ in Eq.~\eqref{eq:threeparaGamma} is no longer valid.

\begin{figure}[hbt!]
\centering
\begin{subfigure}{0.49\textwidth}
\includegraphics[width=\linewidth]{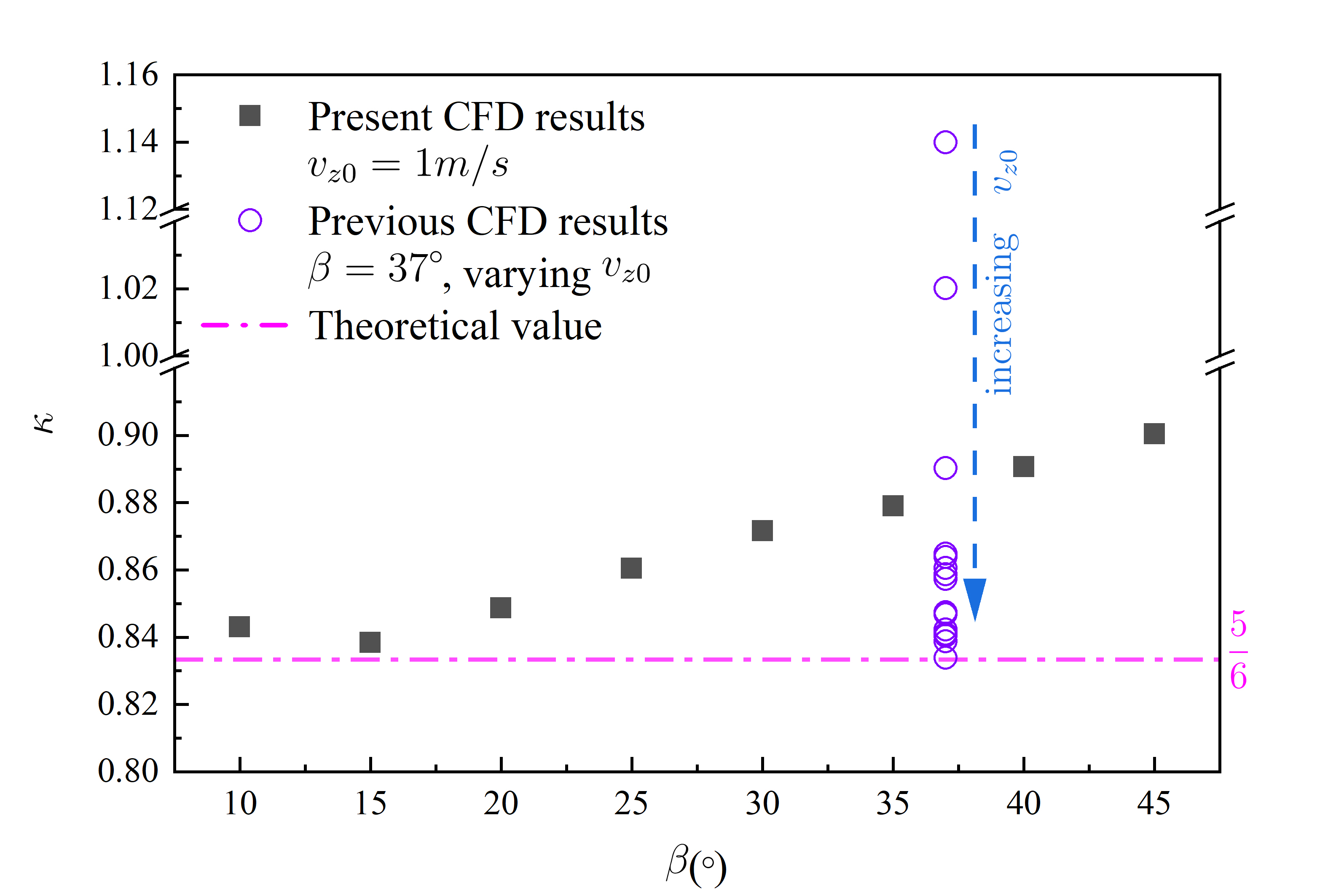} 
\caption{}
\label{fig:DifBetaVz1TZVZKaa}
\end{subfigure}
\begin{subfigure}{0.49\textwidth}
\includegraphics[width=\linewidth]{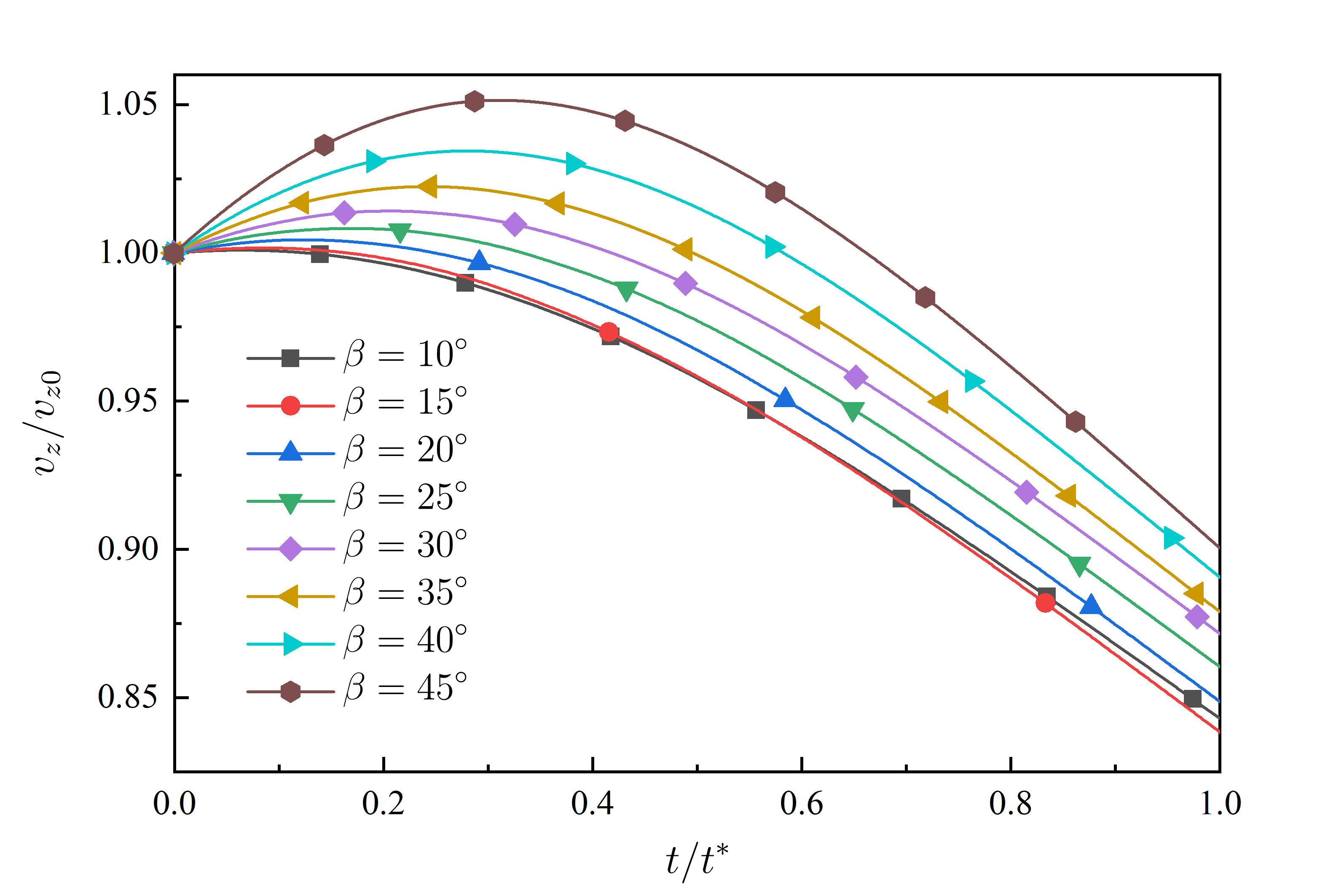}
\caption{}
\label{fig:DifBetaVz1TZVZKab}
\end{subfigure}
\begin{subfigure}{0.49\textwidth}
\includegraphics[width=\linewidth]{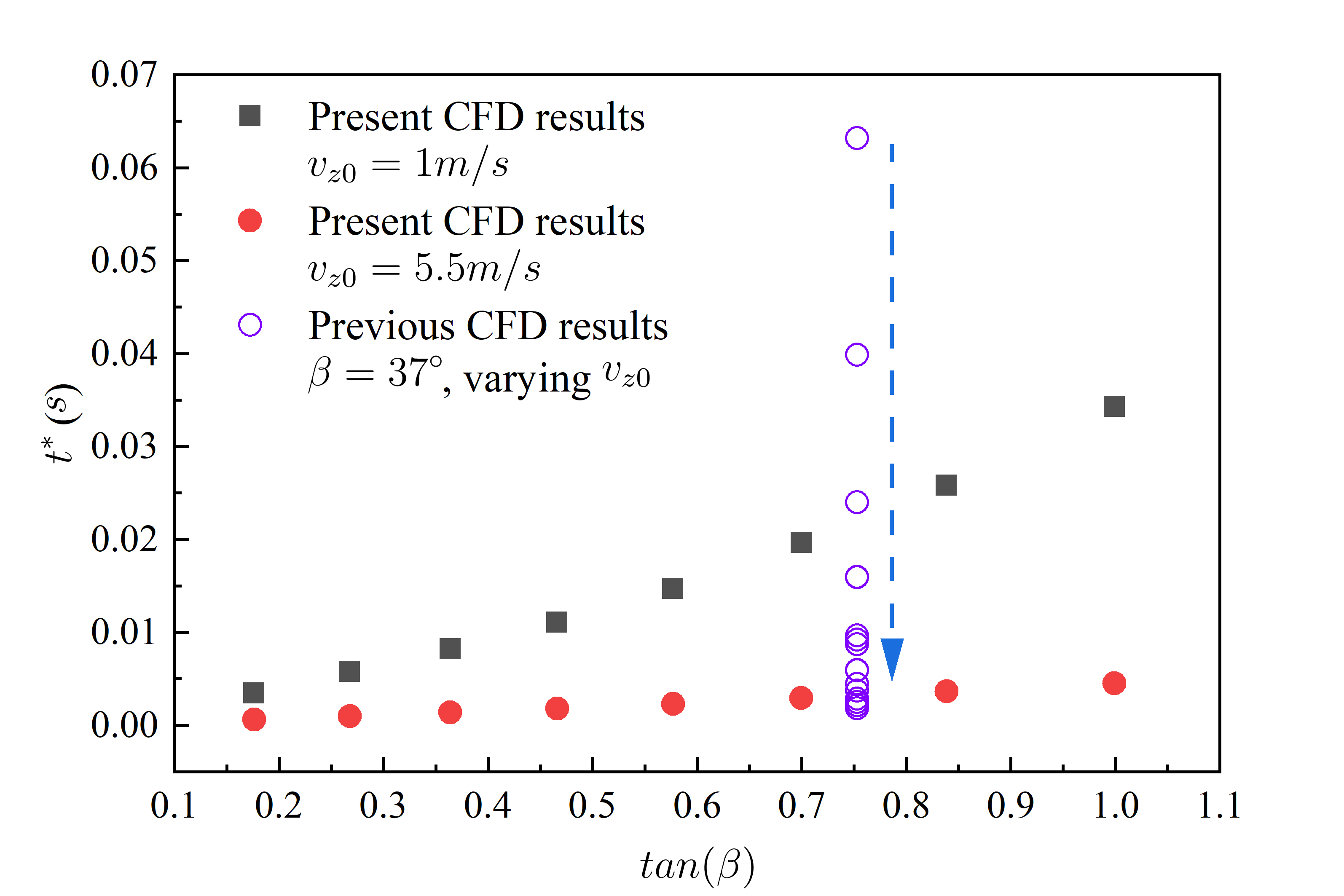} 
\caption{}
\label{fig:DifBetaVz1TZVZKac}
\end{subfigure}
\begin{subfigure}{0.49\textwidth}
\includegraphics[width=\linewidth]{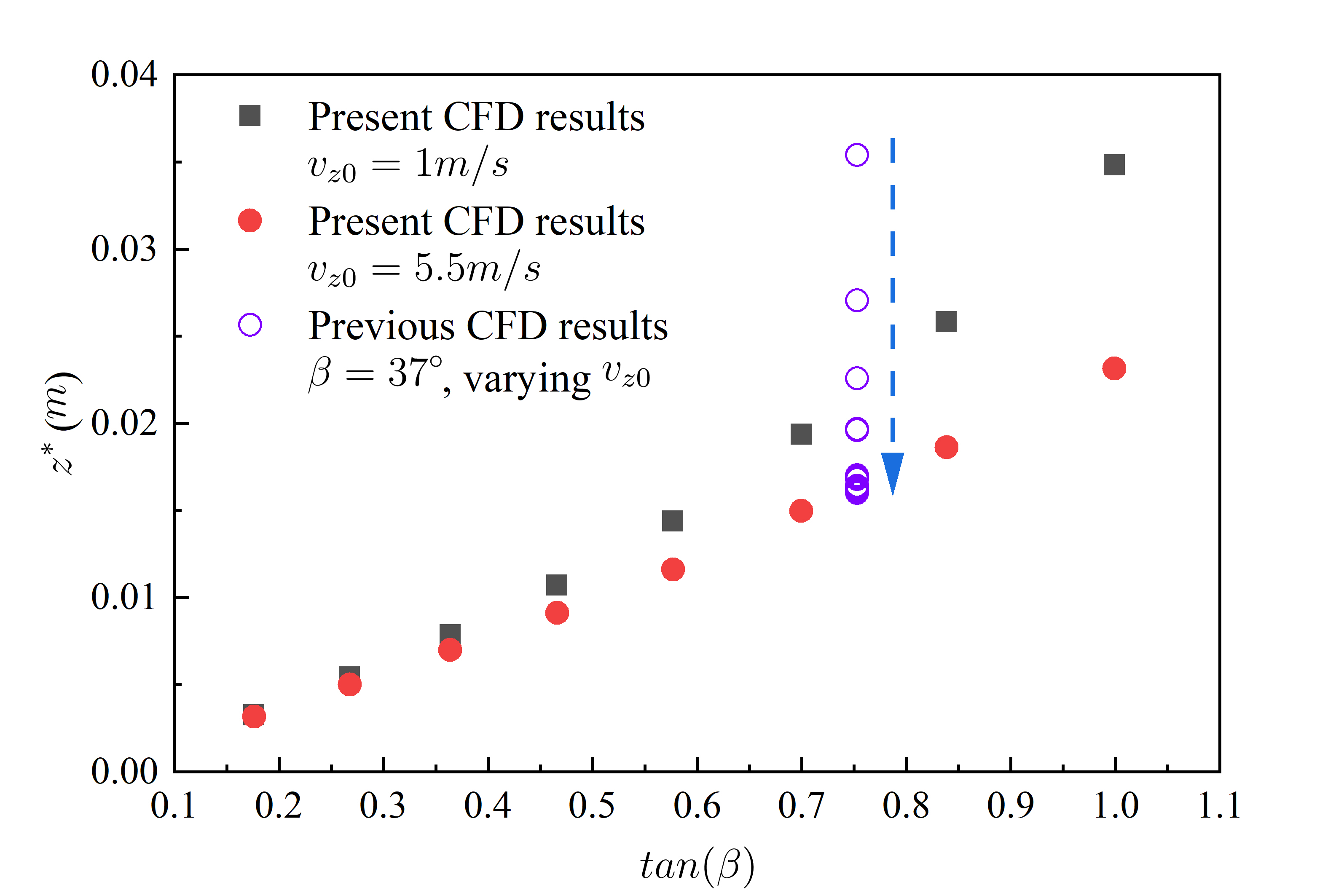}
\caption{}
\label{fig:DifBetaVz1TZVZKad}
\end{subfigure}
\caption{Effect of deadrise angle on dynamic parameters in the case of initial velocity 1 m/s compared with the case of high initial velocity $\upsilon_{z0}$=5.5 m/s and the previous results in the case of $\beta=37^\circ$: (a) $\kappa$; (b) normalized velocity; (c) $t^*$; (d) $z^*$.}
\label{fig:DifBetaVz1TZVZKa}
\end{figure}

Fig.~\ref{fig:DifBetaVz1TZVZKa} shows the comparison between the cases of 1 and 5.5 m/s, respectively. As it can be seen in Fig.~\ref{fig:DifBetaVz1TZVZKaa}, the value of $\kappa$ gradually moves away from the theoretical estimate when the deadrise angle $\beta$ increases, whereas in the case of $\upsilon_{z0}$=5.5 m/s (see Fig.~\ref{fig:DifBetaVz5p5TZVZKab}), a slight deviation is shown between numerical results and theoretical estimate. Moving to Fig.~\ref{fig:DifBetaVz1TZVZKab}, one can see that the accelerating phase becomes more dominant when $\beta$ increases, resulting in the increase of $\kappa$. It indicates that the gravity effect plays a significant role with a large $\beta$ and small initial velocities. On the other hand, when increasing the initial velocity, the gravity effect caused by the deadrise angle can be reduced as shown in Fig.~\ref{fig:DifBetaVz1TZVZKaa}, where the data of $\kappa$ (purple circle) approach the theoretical line gradually for the case of $\beta$=37$^\circ$. The same approaching trends can be also observed on the relationship of $t^*$ and $z^*$ with respect to $tan(\beta)$ (see Fig.~\ref{fig:DifBetaVz1TZVZKac} and Fig.~\ref{fig:DifBetaVz1TZVZKad}). In addition, the linear function can be found although the data obtained from the case of $\upsilon_{z0}$=1 m/s are quite different from the case of $\upsilon_{z0}$=5.5 m/s.

\subsection{Effect of mass}
\label{sec:EffectMass}

\begin{figure}[hbt!]
\centering
\begin{subfigure}{0.49\textwidth}
\includegraphics[width=\linewidth]{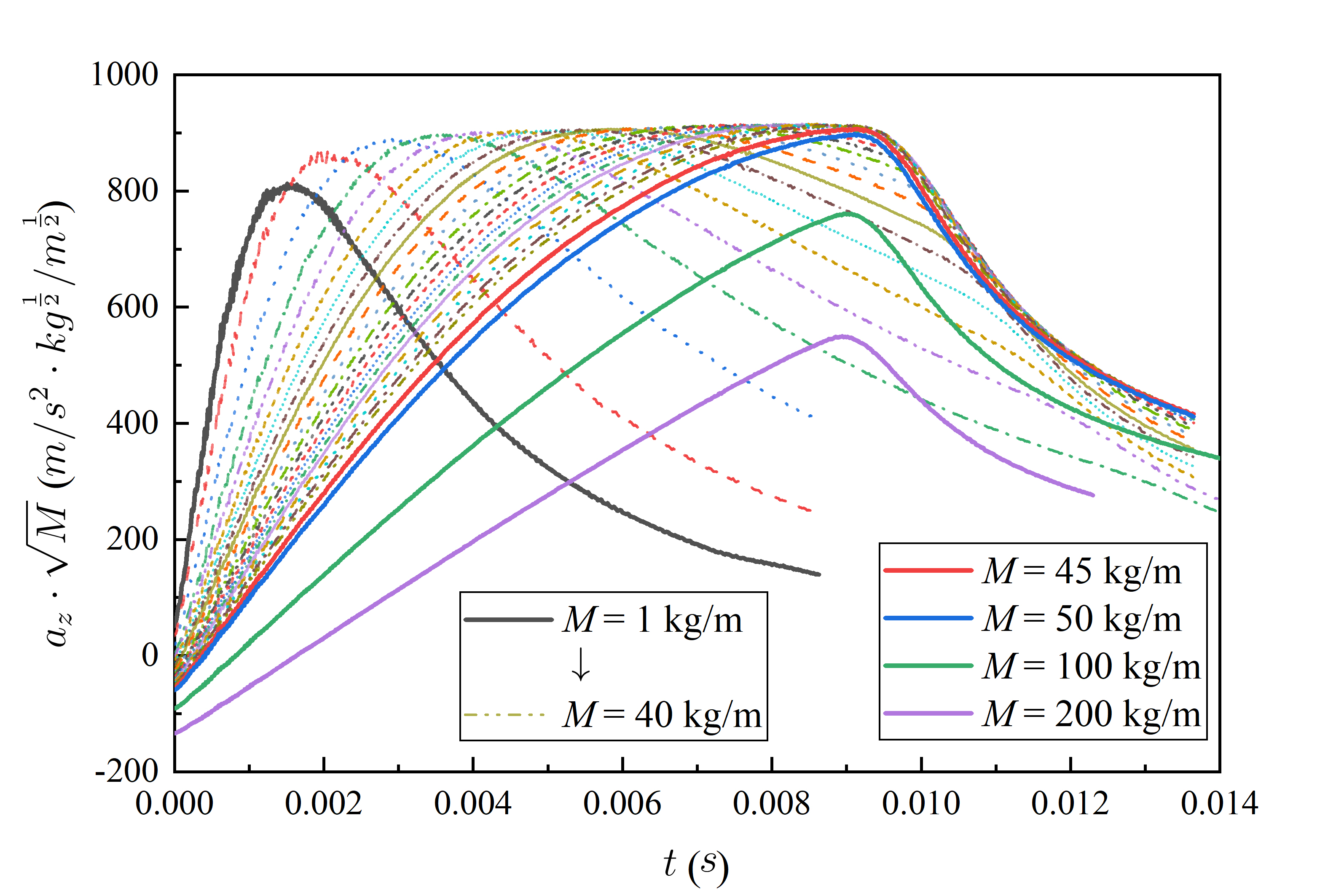}
\caption{}
\label{fig:DifMAzmaxa}
\end{subfigure}
\begin{subfigure}{0.49\textwidth}
\includegraphics[width=\linewidth]{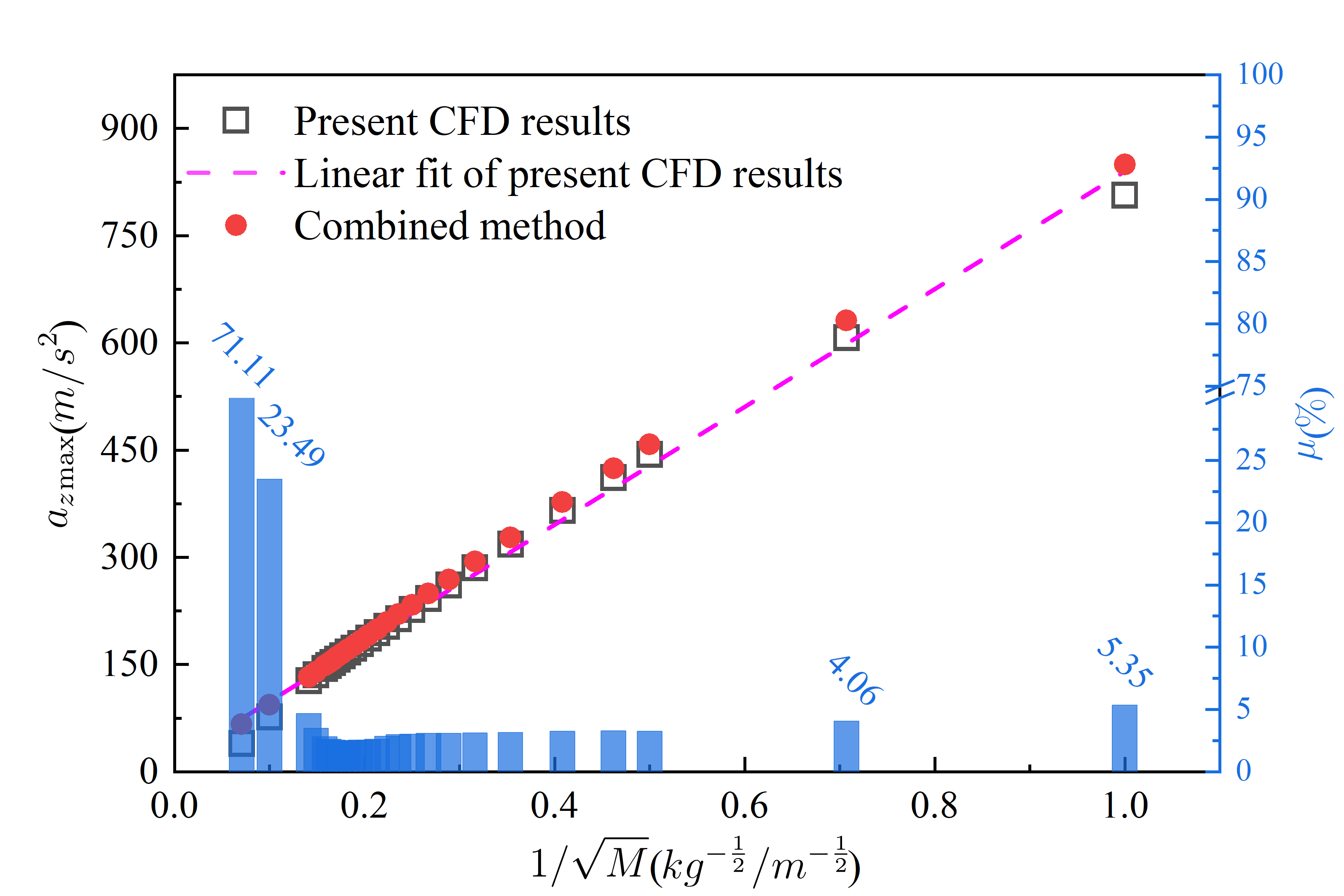}
\caption{}
\label{fig:DifMAzmaxb}
\end{subfigure}
\caption{Effect of mass on maximal acceleration $a_{z\mathrm{max}}$ and the comparison with the results derived from the combined method.}
\label{fig:DifMAzmax}
\end{figure}

In this section, the effect of the variable $M$ on the quantitative relations in Eq.~\eqref{eq:threeparaGamma} is discussed. The mass of the wedge varies from 2 kg/m to 40 kg/m by 2 kg/m, together with additional values, 1 kg/m, 45 kg/m, 50 kg/m, 100 kg/m and 200 kg/m. Note that the initial velocity $\upsilon_{z0}$ of the wedge is 5.5 m/s, associated with the deadrise angle $\beta$~=37$^\circ$: according to the investigation described in the previous section, the pile-up coefficient is chosen as $\gamma_\mathrm{D}$ = 1.14 \citep{iafrati2000hydroelastic,dobrovol1969on}. Following Eq.~\eqref{eq:threeparaGammaA}, for a given deadrise angle and initial velocity of the impacting wedge, the term $a_{z\mathrm{max}} \cdot \sqrt{M}$ should keep constant when the mass changes, but it seems not valid for the case of small and large mass in Fig.~\ref{fig:DifMAzmaxa}. Moreover, Fig.~\ref{fig:DifMAzmaxb} shows the maximal acceleration $a_{z\mathrm{max}}$ computed by the theoretical estimates in Eq.~\eqref{eq:threeparaGammaA} (named as the combined method) compared with the present CFD results. A linear function between $a_{z\mathrm{max}}$ and $1/\sqrt{M}$ can be found among the CFD results which is consistent with \textcolor{black}{the expression in Eq.~\eqref{eq:threeparaGammaA}.} Focusing on the comparison, it can be seen that the \textcolor{black}{theoretical estimate agrees well} with the CFD result with a minor deviation $\mu$ being below 5$\%$ in the range $M \in [2, 40] \mathrm{kg/m}$, except for the case of small and large mass.
In the case of small mass the difference between the theoretical prediction and CFD value can be explained by gravity effect which is not considered in the theoretical approach, but becomes more significant for the wedge impacting event with a relatively small mass.

\begin{figure}[hbt!]
\centering
\includegraphics[width=0.9\textwidth]{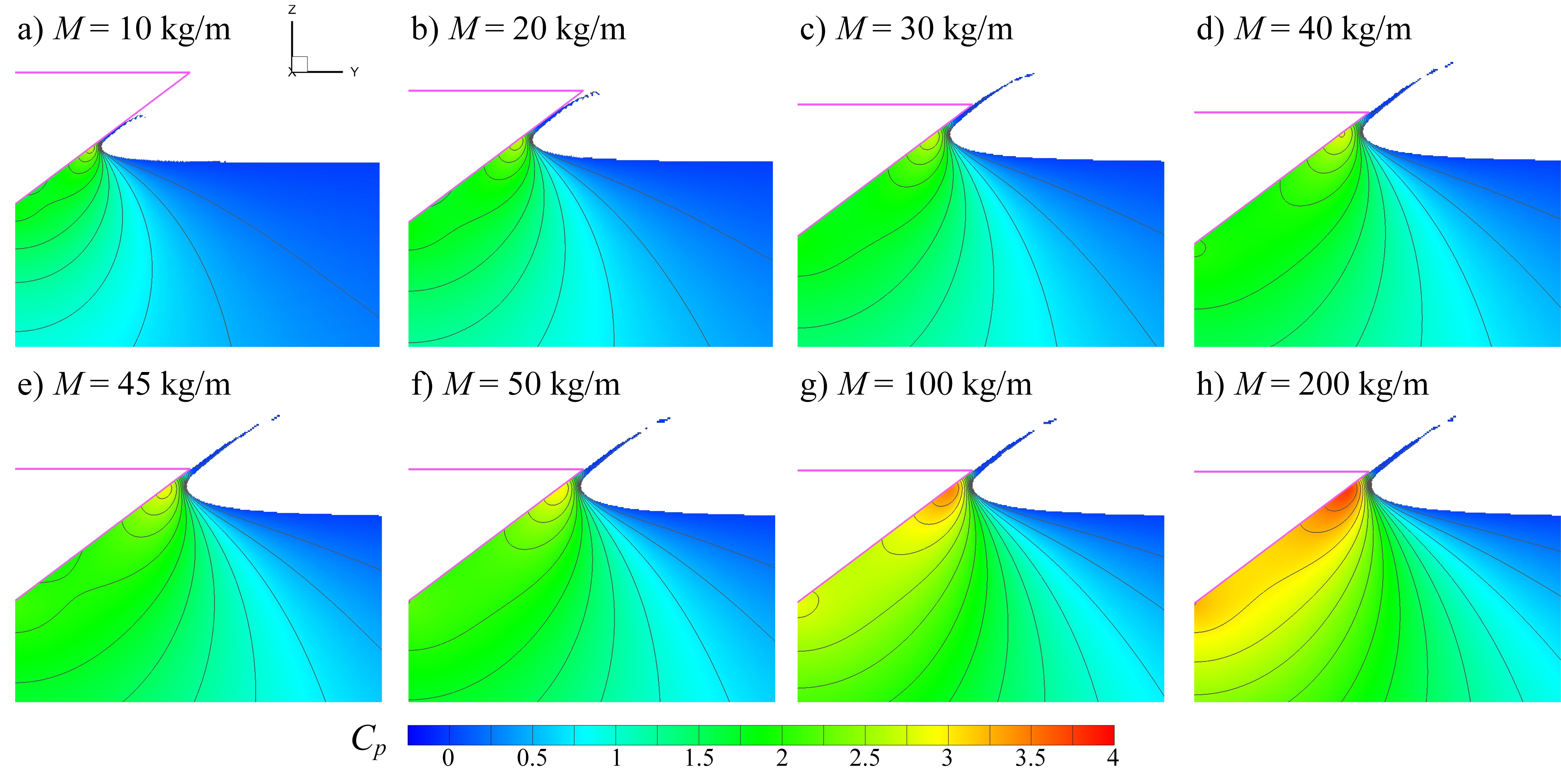}
\caption{Pressure distribution for different mass of wedge when $a_z$ reaches its maximum.}
\label{fig:DifMCp}
\end{figure}

\begin{figure}[hbt!]
\centering
\includegraphics[width=0.49\textwidth]{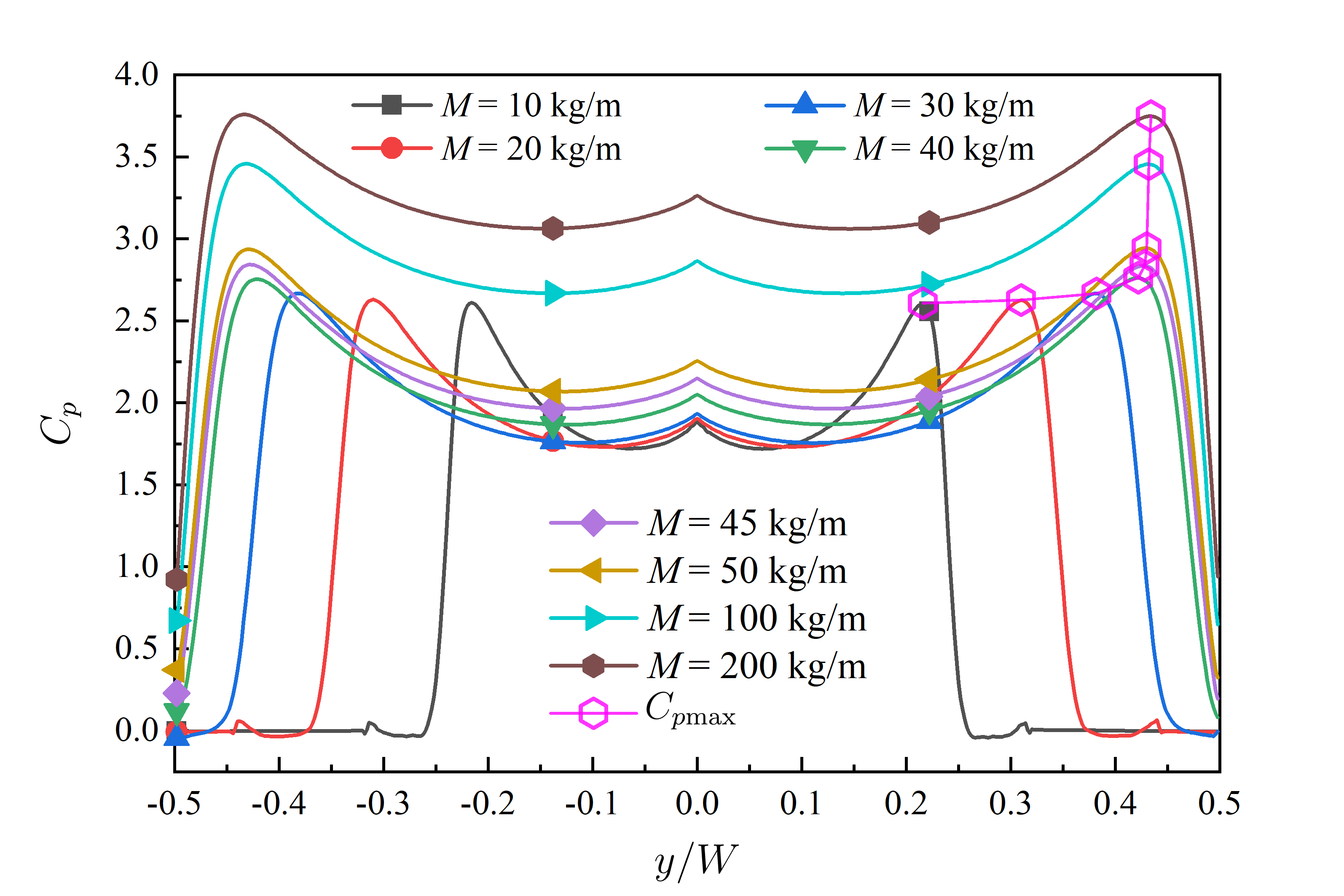}
\caption{Pressure coefficient along the non-dimensional y-axis for different mass of wedge when $a_z$ reaches its maximum.}
\label{fig:DifMCpline}
\end{figure}

\begin{figure}[hbt!]
\centering
\begin{subfigure}{0.49\textwidth}
\includegraphics[width=\linewidth]{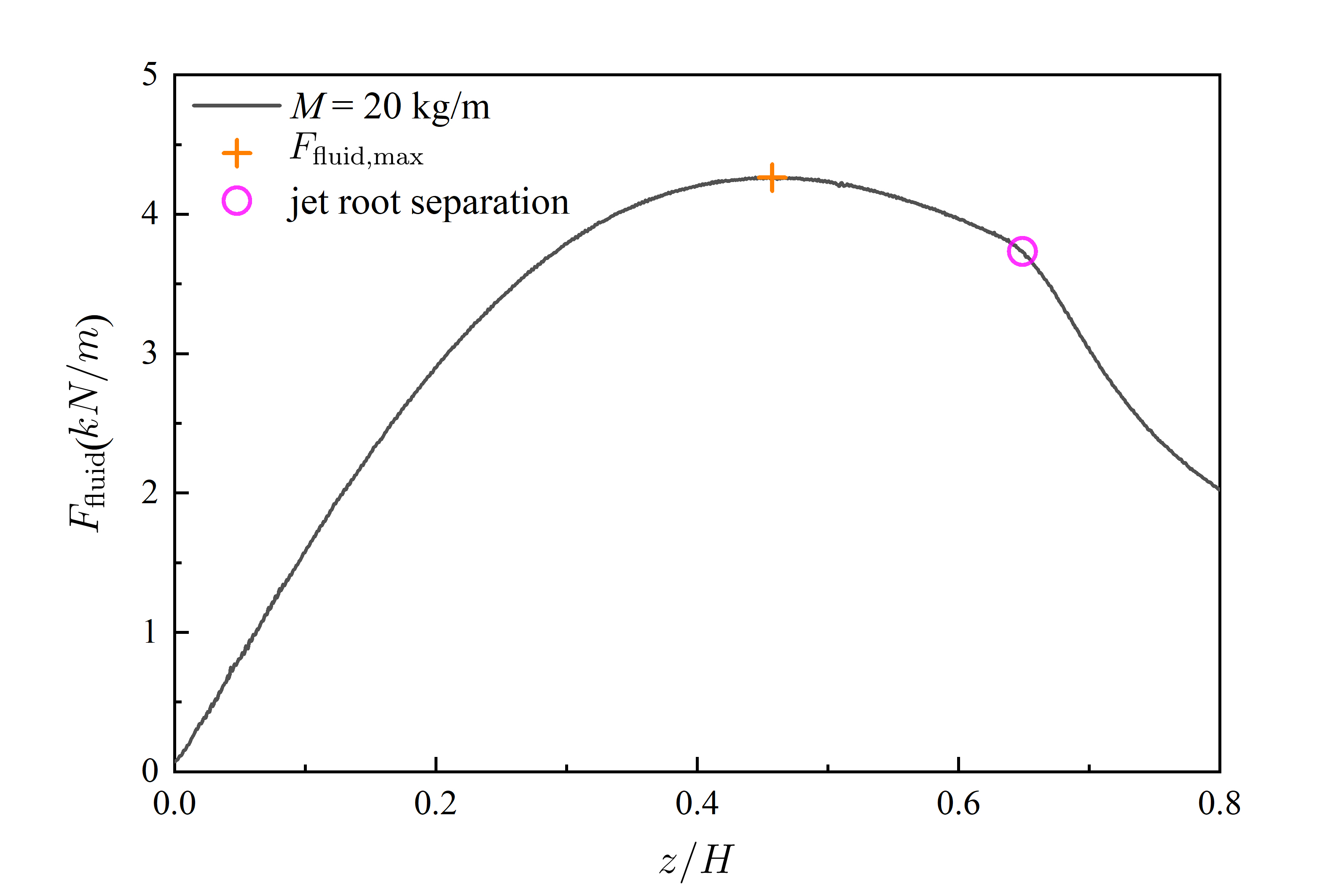}
\caption{}
\label{fig:DifMFfluida}
\end{subfigure}
\begin{subfigure}{0.49\textwidth}
\includegraphics[width=\linewidth]{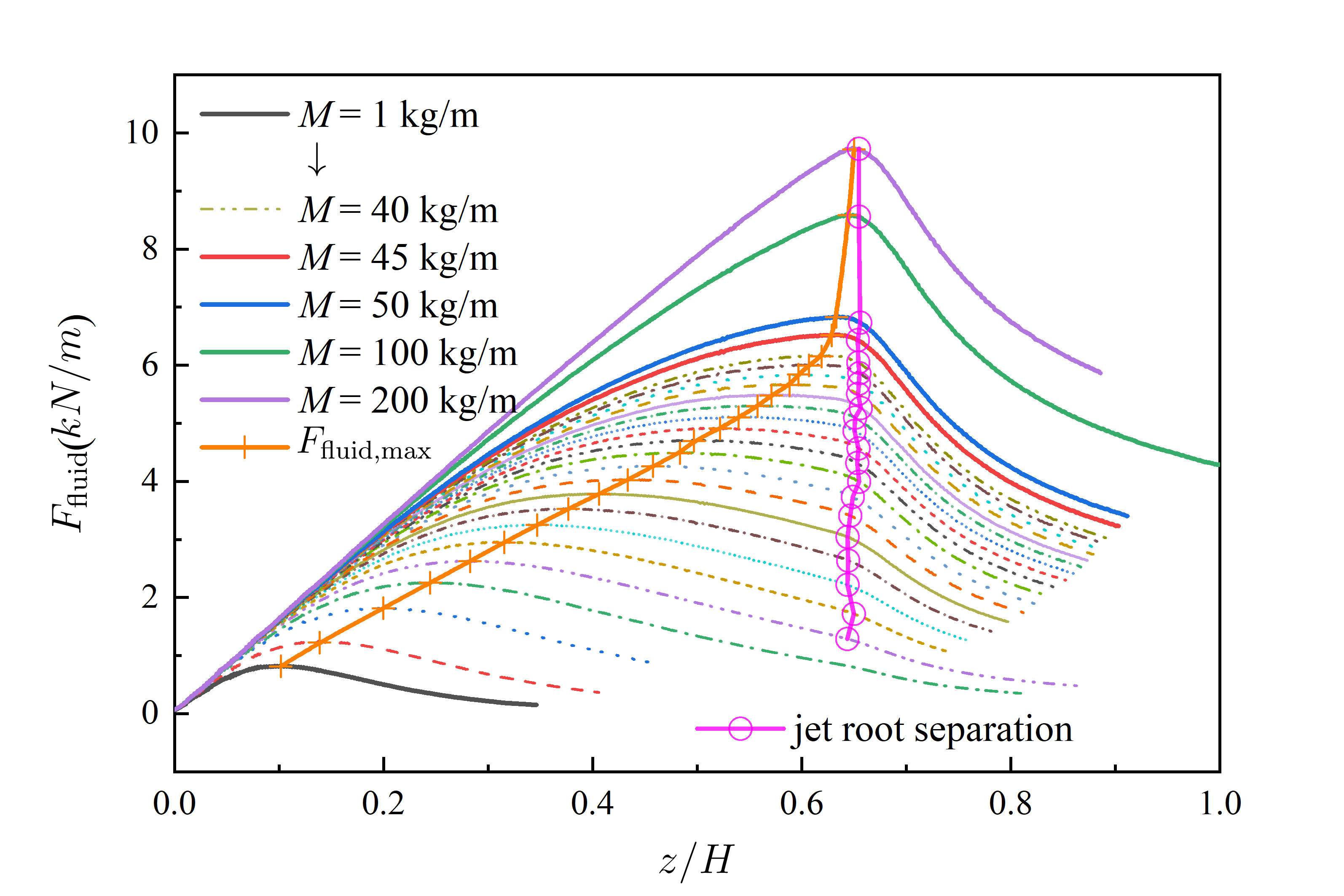}
\caption{}
\label{fig:DifMFfluidb}
\end{subfigure}
\caption{\textcolor{black}{Fluid force acting on the wedge with force peak and the instant of jet root separation during the impact event: (a) the case of $M$=20 kg/m; (b) different mass of the wedge.}}
\label{fig:DifMFfluid}
\end{figure}

Furthermore, in order to understand the difference between the theoretical prediction and CFD results in the case of large mass, the hydrodynamic behaviour in terms of free surface, pressure and force has been investigated.
Fig.~\ref{fig:DifMCp} and Fig.~\ref{fig:DifMCpline} show the water surface elevation and the pressure distribution around the wedge and along the wedge surface when $a_z$ reaches its maximum. The pressure coefficient $C_p$ is defined as:

\begin{equation}
\label{eq:Cp}
C_p=\dfrac{p-p_0}{0.5 \rho \upsilon_{z0}^2}
\end{equation}
It can be observed that with the increasing mass, the pressure peak gradually moves along the wedge surface, and when $M > 40 \mathrm{kg/m}$, since the jet root separated from the chine, the position of the pressure peak approaches a constant value (see magenta points in Fig.~\ref{fig:DifMCpline}). Moreover, the variation of fluid force, $F_\mathrm{fluid}$, with the scaled penetration depth, $z/H$, where $H$ denotes the height of the wedge, is shown in  Fig.~\ref{fig:DifMFfluida} for the case of $M = 20 \mathrm{kg/m}$. One can see that after force peak, which is marked by `+', the force displays firstly a smoothly decreasing trend, and then it drops sharply as the jet root leaves the chine causing a rapid reduction of the pressure in the entire body surface. Thus, the instant in which the jet root is almost separated is approximately tracked by the sharp decreasing of fluid force, denoted as `$\circ$'. It is worth noting that \textcolor{black}{by varying mass, (see Fig.~\ref{fig:DifMFfluidb})}, the instant of the jet root separation occurrence is almost a constant value ($z/H \approx 0.65$) that is similar to the constant velocity water-entry \citep{wen2022formualtions}. Moreover, by increasing mass the instant at which force reaches peak progressively coincides with the instant when the jet root is almost separated. Hence, considering that the body acceleration is strictly dependent on the fluid force in a free fall impact ($a_z = (F_\mathrm{fluid} - Mg)/M$), also the maximum acceleration value is ``forced" by the separation of the jet root, and this is why the theoretical prediction, which doesn't consider the separation phenomena, overestimates the $a_{z\mathrm{max}}$ (see Fig.~\ref{fig:DifMAzmaxb}).

\begin{figure}[hbt!]
\centering
\begin{subfigure}{0.49\textwidth}
\includegraphics[width=\linewidth]{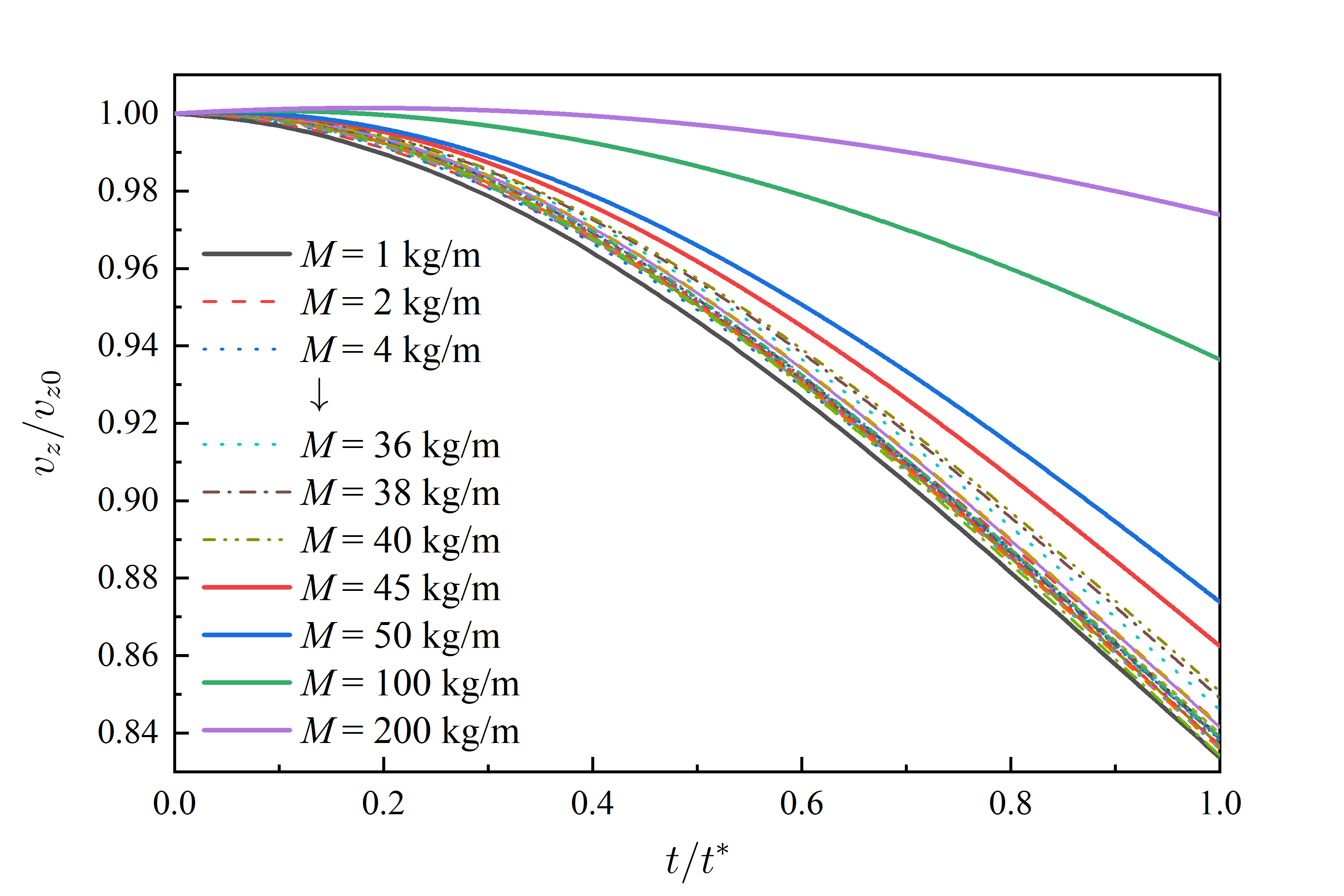}
\caption{}
\label{fig:DifMVzTZVZKaa}
\end{subfigure}
\begin{subfigure}{0.49\textwidth}
\includegraphics[width=\linewidth]{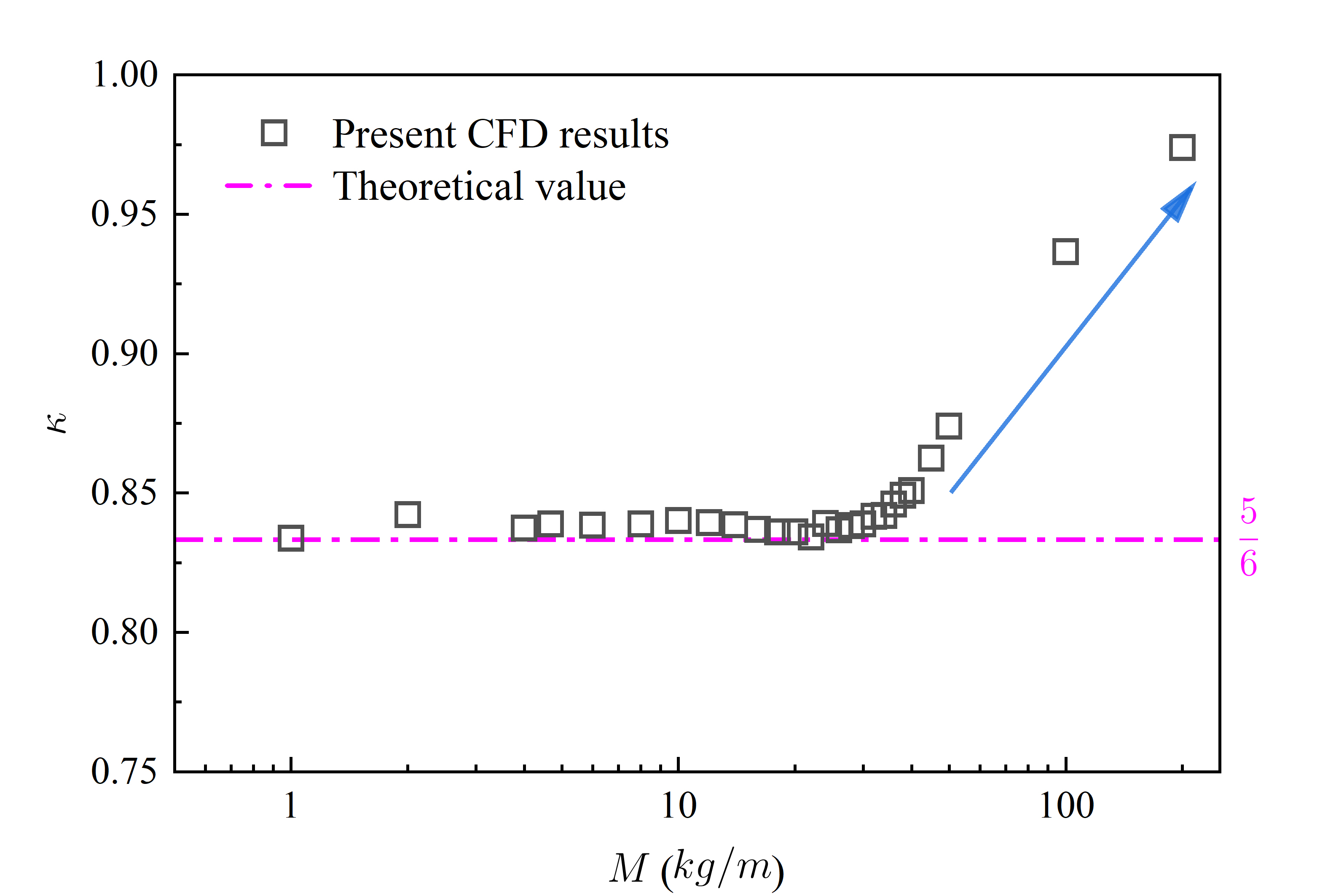}
\caption{}
\label{fig:DifMVzTZVZKab}
\end{subfigure}
\begin{subfigure}{0.49\textwidth}
\includegraphics[width=\linewidth]{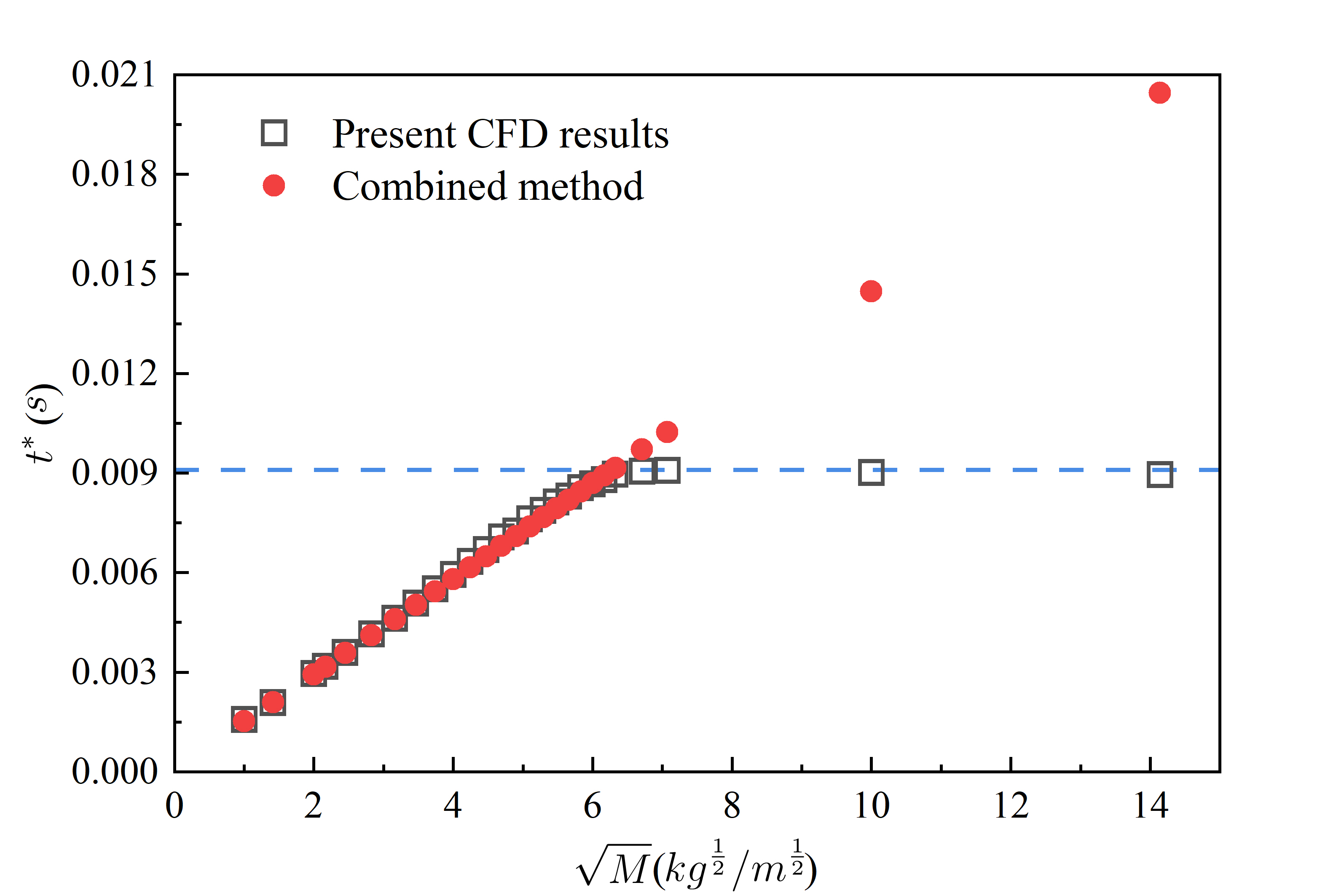} 
\caption{}
\label{fig:DifMVzTZVZKac}
\end{subfigure}
\begin{subfigure}{0.49\textwidth}
\includegraphics[width=\linewidth]{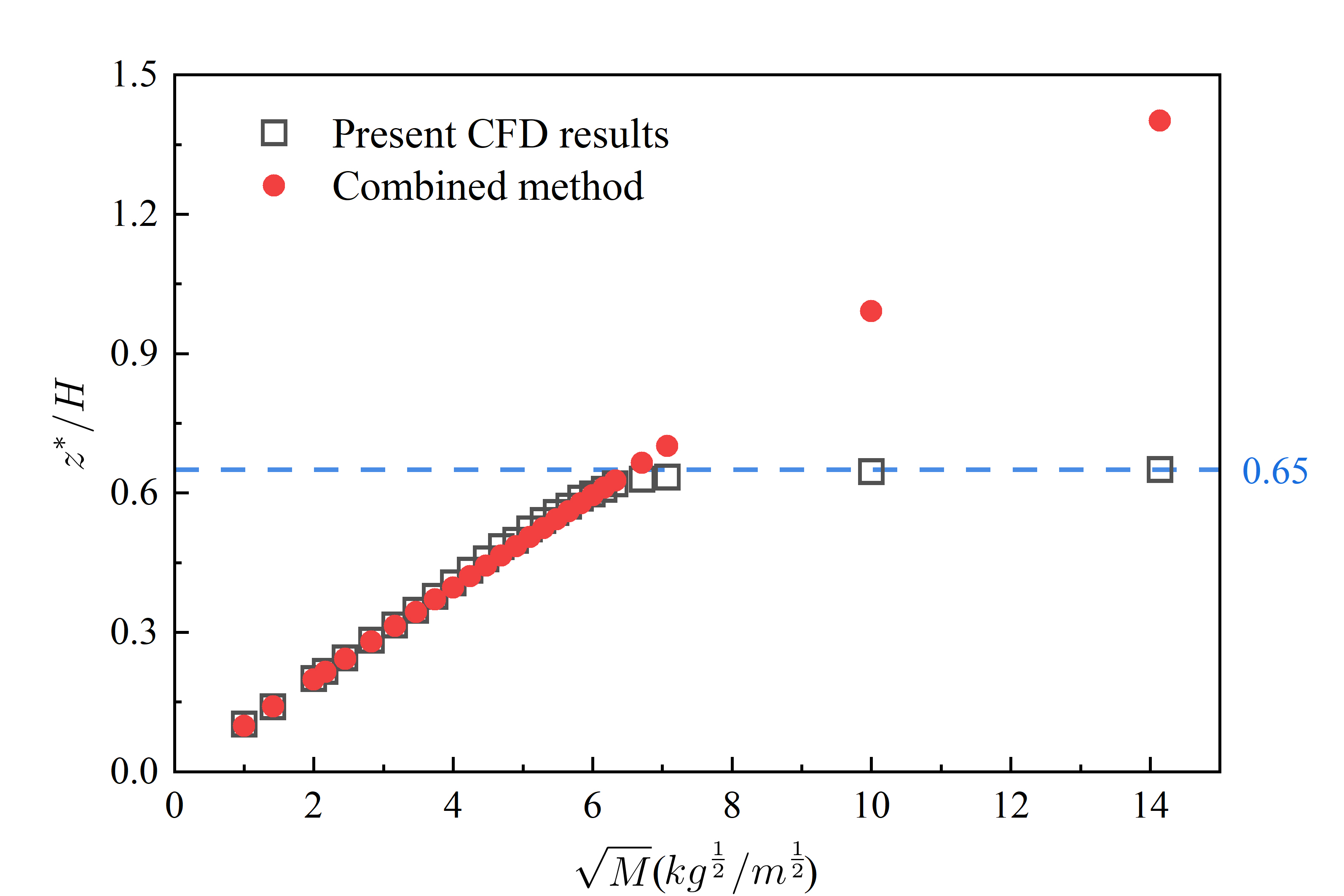}
\caption{}
\label{fig:DifMVzTZVZKad}
\end{subfigure}
\caption{Effect of mass on variable dynamic parameters in the case of initial velocity 5.5 m/s: (a) $t^*$; (b) $z^*$; (c) $\upsilon_z^*$; (d) $\kappa$ \textcolor{black}{(bule dashed line in (c) and (d) indicates the instant of the jet root separation occurrence)}.}
\label{fig:DifMVzTZVZKa}
\end{figure}

In order to further examine the combined method, referring to Eq.~\eqref{eq:threeparaGamma}, the value of the ratio of velocity $\kappa$, time $t^*$ and penetration depth $z^*$ at which $a_z$ reaches its maximal value are provided in Fig.~\ref{fig:DifMVzTZVZKa}. It is worth noticing that a good collapse of the non-dimensional velocity versus the dimensionless time ($\upsilon_z/\upsilon_{z0}\sim t/t^*$) is found when $M \in [2, 40] \mathrm{kg/m}$, meaning that, during the impact, the variation of velocity is independent of the mass of wedge and the gravity effect is not so important in these cases with a large initial velocity. As it is shown in Fig.~\ref{fig:DifMVzTZVZKab}, the parameter $\kappa$ computed numerically is close to the theoretical estimate 5/6 with a root mean square error (RMSE) of 0.005117. However, when the mass increases beyond 40 kg/m, $\kappa$ moves far away from the theoretical value 5/6 as it is also confirmed by Fig.~\ref{fig:DifMAzmaxb}. Moving to the relations of $t^*$ and $z^*$ with respect to $\sqrt{M}$ in Fig.~\ref{fig:DifMVzTZVZKac} and ~\ref{fig:DifMVzTZVZKad}, when the mass is less than or equal to 40 kg/m, the linear trends are displayed by numerical approach with various square root of mass $\sqrt{M}$ that is consistent with the theoretical relation shown in Eq.~\eqref{eq:threeparaGammaT} and Eq.~\eqref{eq:threeparaGammaZ}. So, also for these variables, the results derived from the combined method have a good agreement with the numerical results, as observed on the prediction of $a_{z\mathrm{max}}$ in Fig.~\ref{fig:DifMAzmaxb}. Besides, since the jet root is almost completely separated from the wedge for $M > 40 \mathrm{kg/m}$, the numerical results of $t^*$ and $z^*$ become constant and progressively diverge from theoretical predictions in which the separation phenomenon is not considered.

\color{black}

\subsection{Validity of the theoretical equations}
\label{sec:Validity}

\begin{figure}[hbt!]
\centering
\begin{subfigure}{0.65\textwidth}
\includegraphics[width=\linewidth]{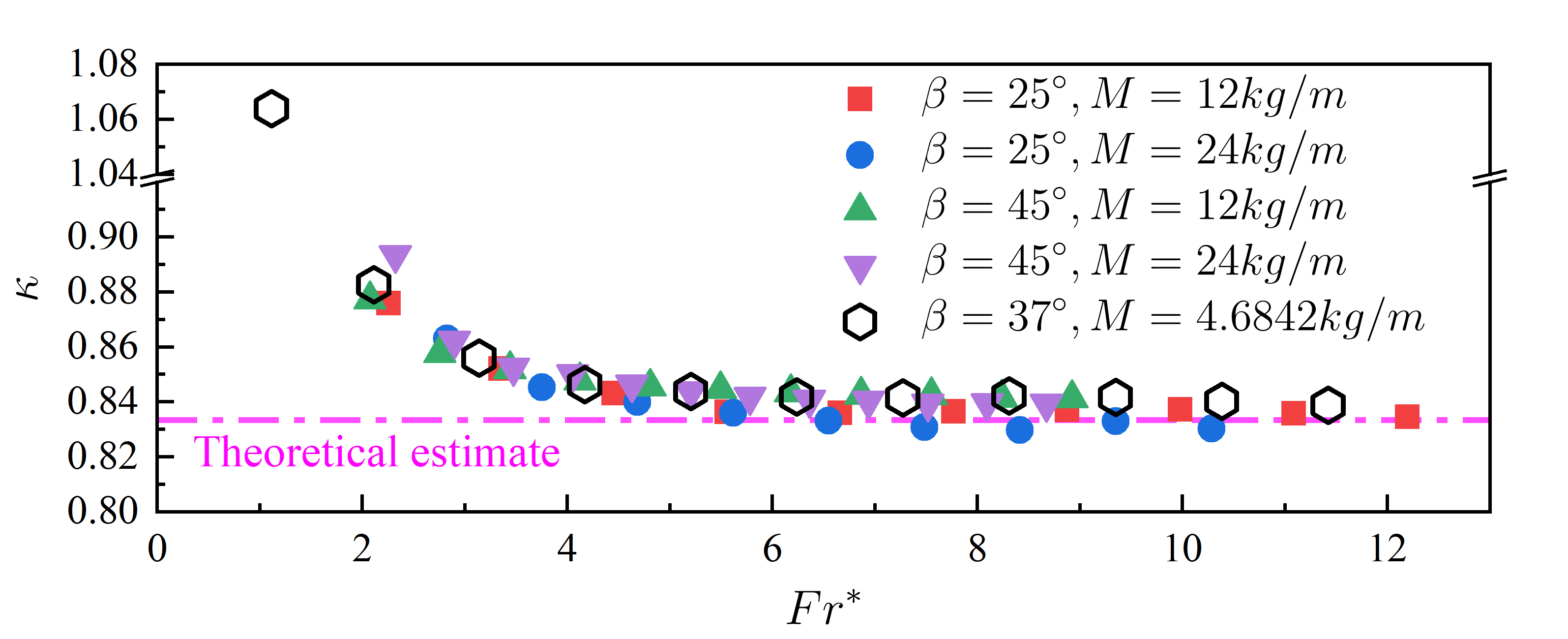}
\caption{}
\label{fig:FrZstarKappaa}
\end{subfigure}
\begin{subfigure}{0.65\textwidth}
\includegraphics[width=\linewidth]{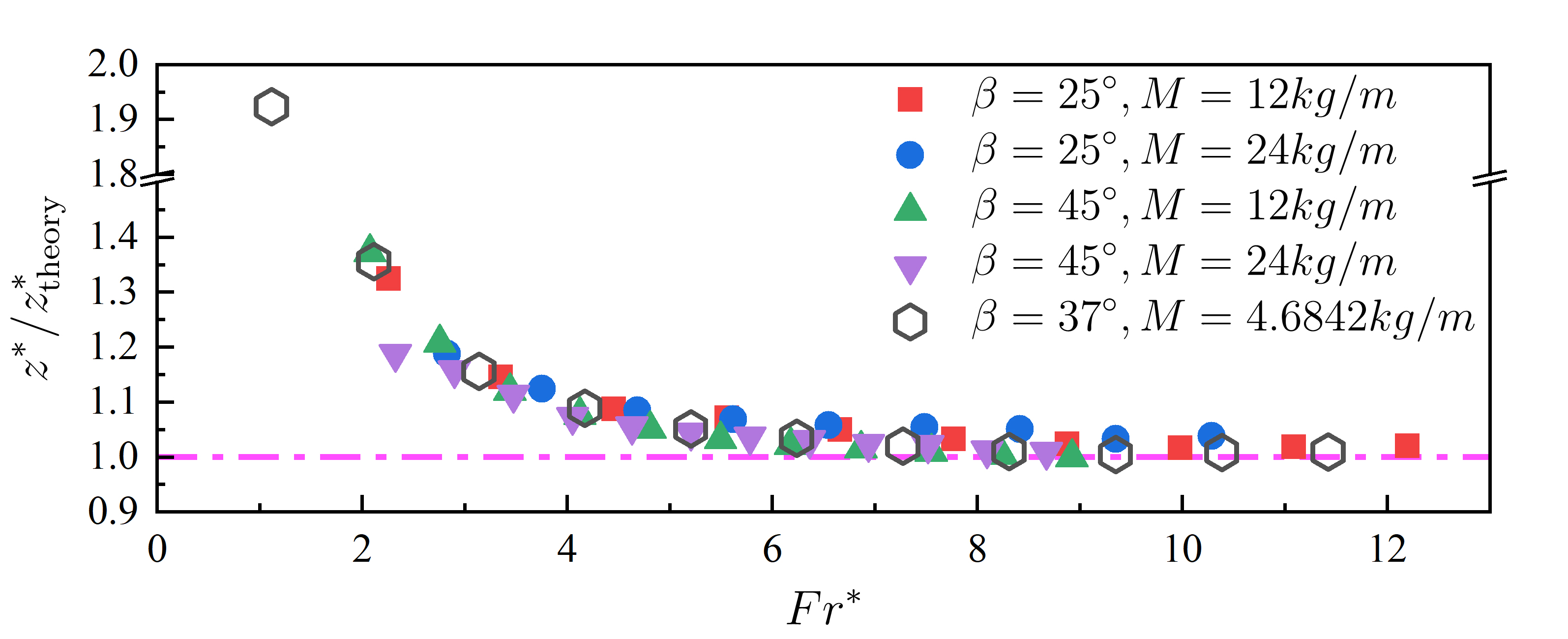} 
\caption{}
\label{fig:FrZstarKappab}
\end{subfigure}
\caption{The results of the four typical conditions associated with the original case when $Fr^*$ changes: (a) the ratio $\kappa$; (b) dimensionless penetration depth $z^*/z^*_\mathrm{theory}$.}
\label{fig:FrZstarKappa}
\end{figure}

The above results prove that it is possible to predict the maximal acceleration and correlated parameters with the help of the combined method, proposed in Eq.~\eqref{eq:threeparaGamma}, with large initial vertical velocity. Thus, it is meaningful to determine the threshold value of initial velocity, below which the combined method is invalid. To this purpose, the instantaneous Froude number $Fr^*$ \citep{hulin2022gravity} is introduced as:

\begin{equation}
\label{eq:Fr}
Fr^* = \dfrac{\upsilon^*_{z\mathrm{theory}}}{\sqrt{gz^*_\mathrm{theory}}}
= \dfrac{5\upsilon_{z0}}{6\sqrt{g\tan(\beta)\sqrt{\dfrac{2M}{5\pi\rho\gamma^2}}}}
\end{equation}
where $\upsilon^*_{z\mathrm{theory}}$ and $z^*_\mathrm{theory}$ are the theoretical estimates referring to Eqs.~\eqref{eq:threeparaGammaV} and \eqref{eq:threeparaGammaZ}. Considering the applicability of the equations (see in Eq.~\eqref{eq:threeparaGamma}), other four groups of typical initial parameters ($\beta, M$) with varying initial velocity $\upsilon_{z0}$ are taken into account, i.e. $\beta=25^\circ, M=12\mathrm{kg/m}; \beta=25^\circ, M=24\mathrm{kg/m}; \beta=45^\circ, M=12\mathrm{kg/m}; \beta=45^\circ, M=24\mathrm{kg/m}$ and the results of these four conditions associated with the case of $\beta=37^\circ, M=4.6842\mathrm{kg/m}$ are presented in Fig.~\ref{fig:FrZstarKappa}. \textcolor{black}{A good} overlapping trends can be observed on both the ratio $\kappa$ and dimensionless penetration depth $z^*/z^*_\mathrm{theory}$ and the results tend to approach the theoretical value with increasing $Fr^*$, indicating the independence of correlated parameters $\kappa$ and $z^*$ on the deadrise angle $\beta$ and the mass $M$.


\begin{figure}[hbt!]
\centering
\begin{subfigure}{0.49\textwidth}
\includegraphics[width=\linewidth]{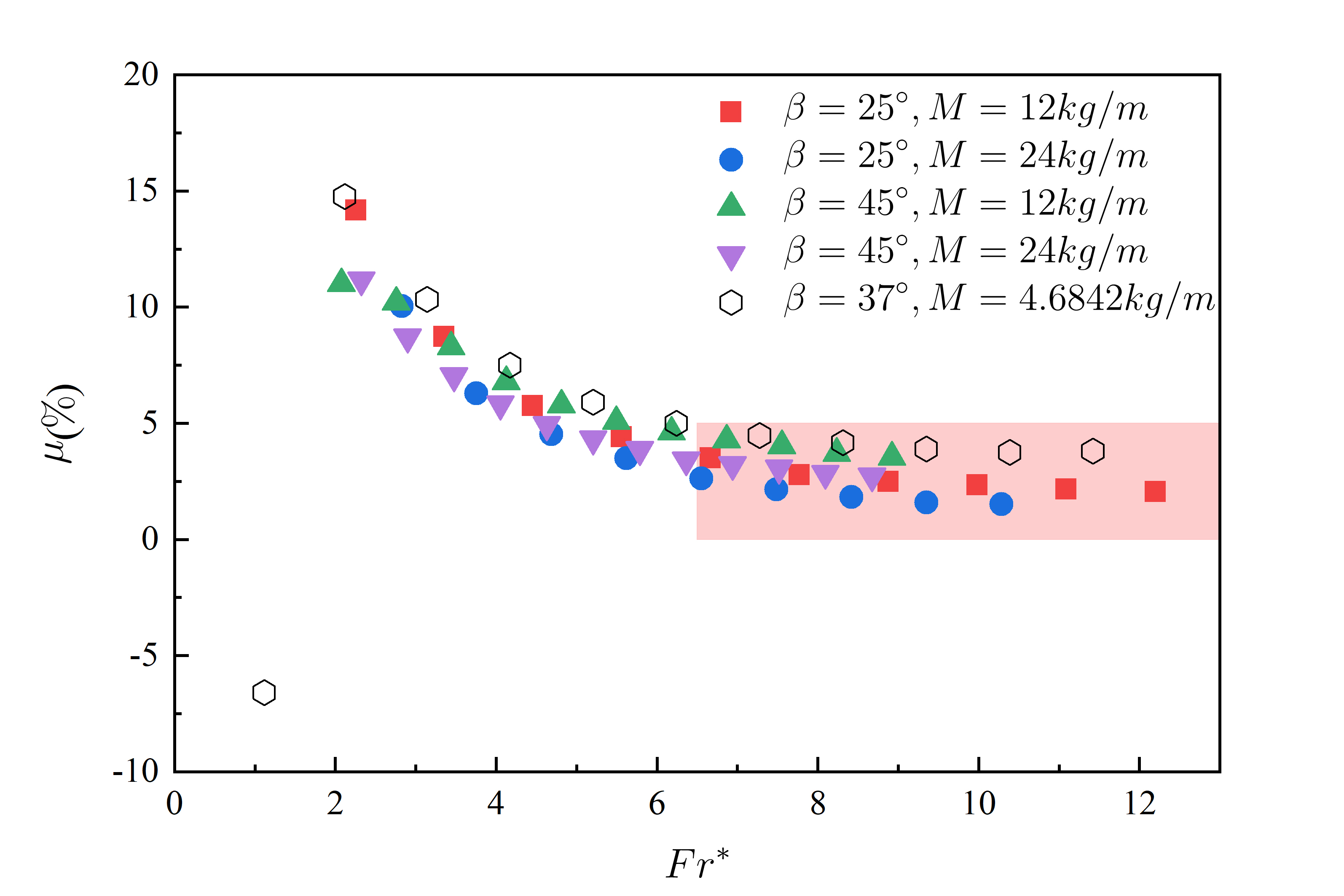}
\caption{}
\label{fig:Compamua}
\end{subfigure}
\begin{subfigure}{0.49\textwidth}
\includegraphics[width=\linewidth]{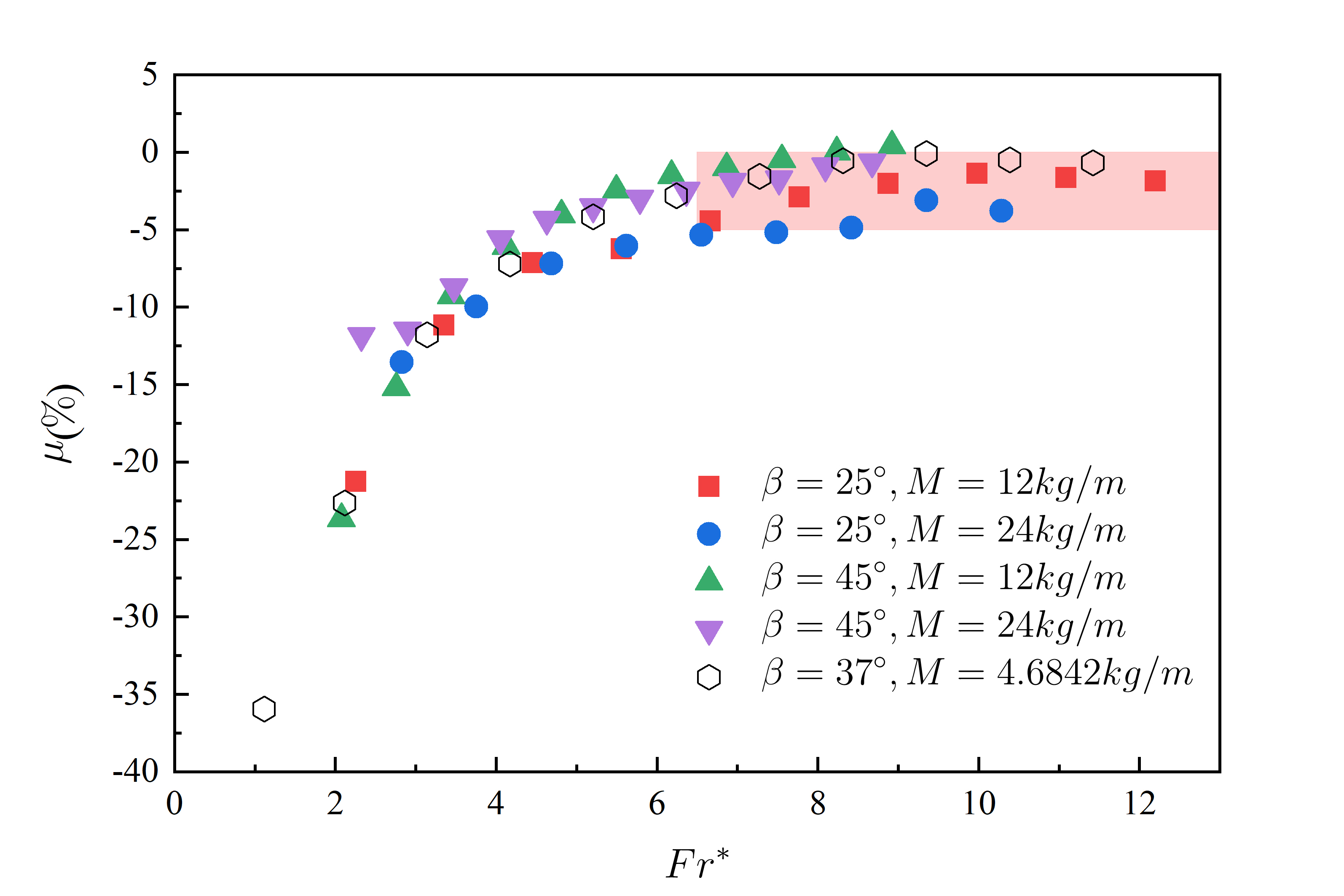} 
\caption{}
\label{fig:Compamub}
\end{subfigure}
\caption{The difference between the theoretical prediction and the numerical result for various $Fr^*$ on: (a) the maximal acceleration $a_{z\mathrm{max}}$; (b) the corresponding time $t^*$.}
\label{fig:Compamu}
\end{figure}

The comparison of the theoretical estimate and numerical result on the maximal acceleration $a_{z\mathrm{max}}$ is presented in Fig.~\ref{fig:Compamua}, where the deviation $\mu$ is defined as $\mu = (a_{z\mathrm{max\_theory}} - a_{z\mathrm{max\_CFD}}) / a_{z\mathrm{max\_CFD}}$. Note that $a_{z\mathrm{max\_theory}}$ is computed by Eq.~\eqref{eq:threeparaGammaA}. One can see that when the instantaneous Froude number $Fr^*$ is greater than 6.5 (see the light pink region), the deviation $\mu$ is below 5$\%$ and decreases with the increase of $Fr^*$. A similar trend can be also observed for the deviation $\mu$ of the corresponding time $t^*$ which shows that the theoretical prediction underestimates the numerical results, as shown in Fig.~\ref{fig:Compamub}. Hence, it seems that the combined method, by using Eq.~\eqref{eq:threeparaGamma}, accurately predicts the maximal acceleration $a_{z\mathrm{max}}$ and the correlated parameters ($z^*$, $v_z^*$ and $t^*$) for a wide range of deadrise angle and mass of a wedge when the instantaneous Froude number $Fr^*$ (see Eq. \eqref{eq:Fr}) is greater than 6.5.

\subsection{Scaling laws on free-falling water-entry}

\begin{figure}[hbt!]
\centering
\includegraphics[width=0.49\textwidth]{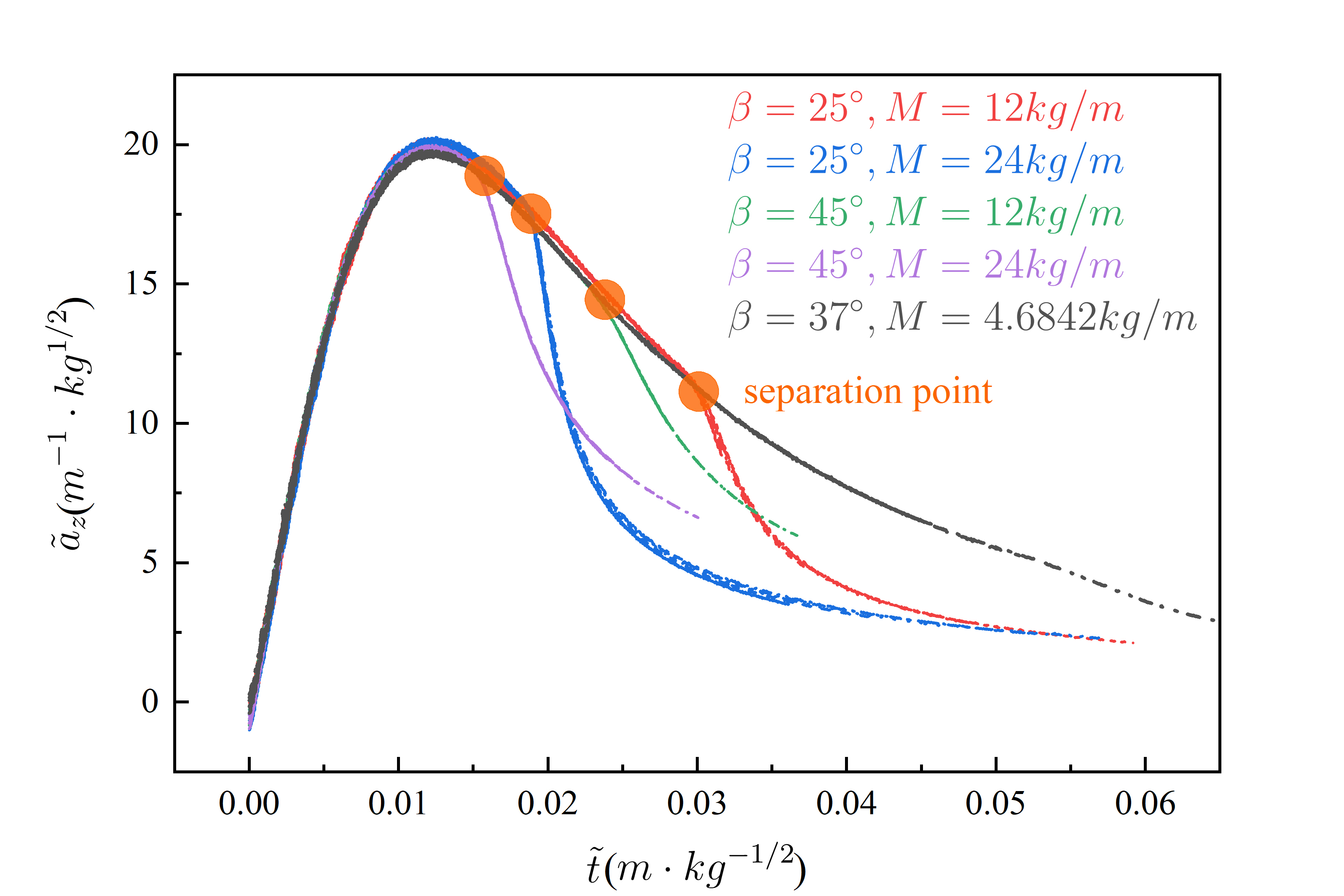}
\caption{Variation of scaled acceleration $\tilde{a}_z$ on scaled time $\tilde{t}$ for all the valid cases on the five conditions \textcolor{black}{(different line type for each set of initial parameters $(\beta, M)$ distinguishes the cases with different initial velocity)}.}
\label{fig:Scaled_AzT}
\end{figure}

\textcolor{black}{Based on} the above results, it can be concluded that $a_{z\mathrm{max}}$, $\upsilon_z^*$, $z^*$ and $t^*$ are dominated by three key parameters ($\beta$, $M$ and $\upsilon_{z0}$). Inspired by this, the scaled acceleration $\tilde{a}_z$ and the scaled time $\tilde{t}$ is firstly obtained as:

\begin{subequations}
\label{eq:Scaled_azt}
\begin{eqnarray}
\label{eq:Scaled_azt_a}
\tilde{a}_z &=& a_z (t) \cdot \dfrac{\sqrt{M}}{\upsilon_{z0}^2} \cdot \dfrac{\tan(\beta)}{\gamma_\mathrm{D}}\\
\label{eq:Scaled_azt_t}
\tilde{t} &=& t \cdot \dfrac{\upsilon_{z0}}{\sqrt{M}} \cdot \dfrac{\gamma_\mathrm{D}}{\tan(\beta)}
\end{eqnarray}
\end{subequations}
In Fig.~\ref{fig:Scaled_AzT}, it is worth noting that, for the wedge impacting water freely, the overlapping trend on the scaled acceleration, derived from the cases in the light pink region in Fig.~\ref{fig:Compamu}, can also be observed before the jet root separation.

\begin{figure}[hbt!]
\centering
\begin{subfigure}{0.49\textwidth}
\includegraphics[width=\linewidth]{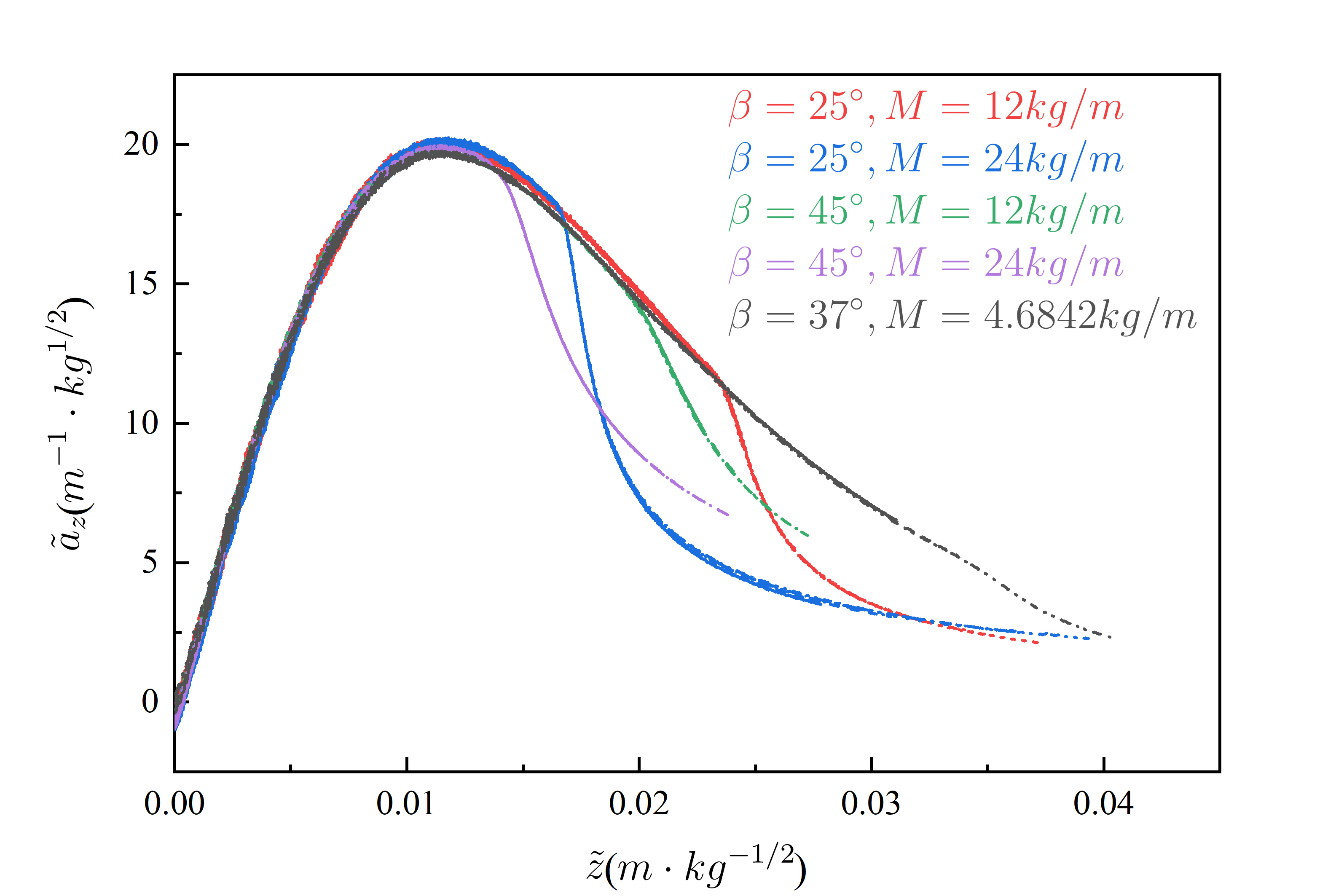}
\caption{}
\label{fig:Scaled_AzVzZa}
\end{subfigure}
\begin{subfigure}{0.49\textwidth}
\includegraphics[width=\linewidth]{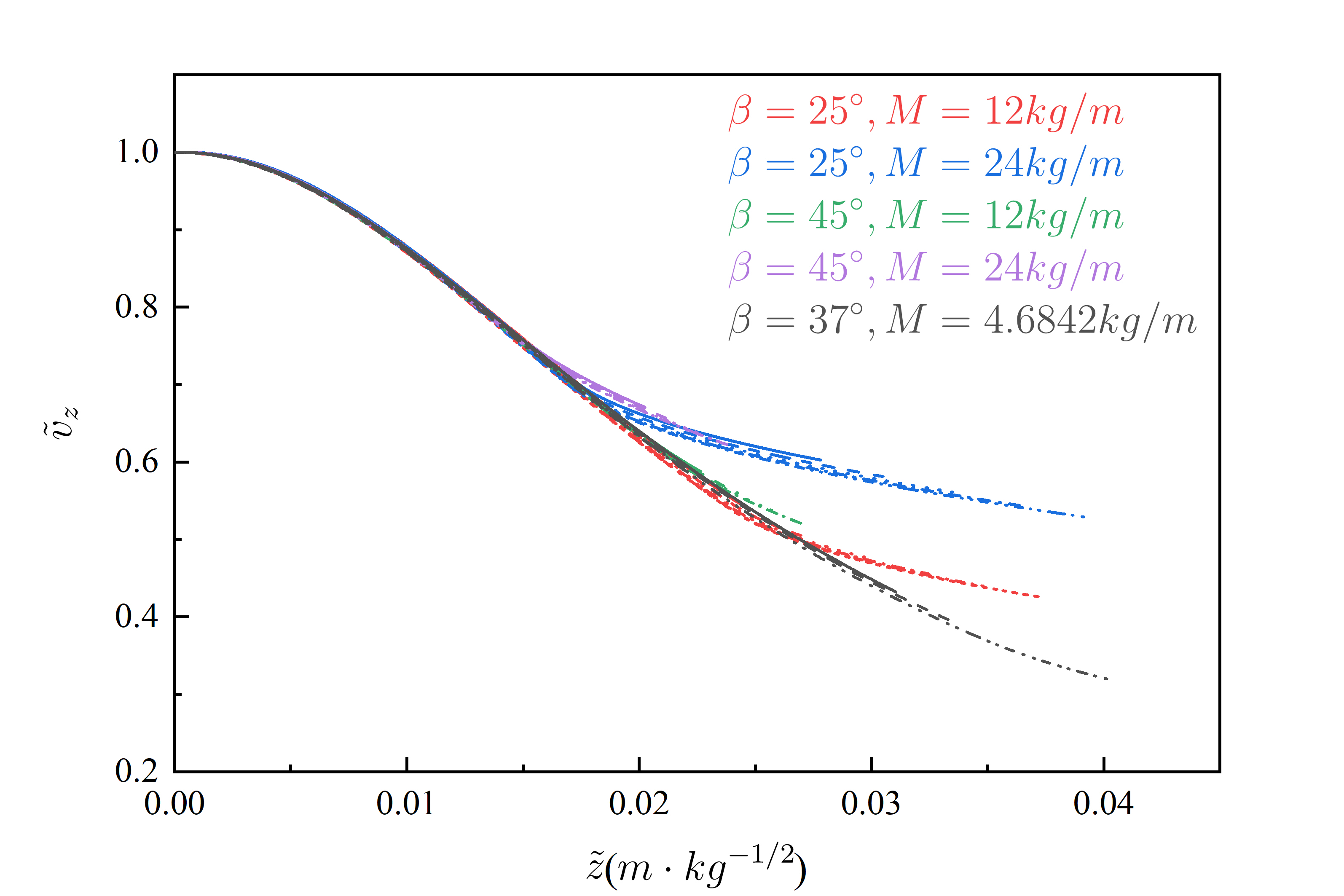} 
\caption{}
\label{fig:Scaled_AzVzZb}
\end{subfigure}
\caption{Variation of scaled values for all the valid cases on the five conditions \textcolor{black}{(different line type for each set of initial parameters $(\beta, M)$ distinguishes the cases with different initial velocity)}: (a) $\tilde{a}_z$ versus $\tilde{z}$; (b) $\tilde{\upsilon}_z$ versus $\tilde{z}$.}
\label{fig:Scaled_AzVzZ}
\end{figure}

Based on the good collapse of the scaled acceleration histories from different initial impacting velocity $\upsilon_{z0}$, deadrise angle $\beta$ and mass $M$, the other two overlapping trends about $\tilde{a}_z$ and $\tilde{\upsilon}_z$ as functions of $\tilde{z}$ are presented in Fig.~\ref{fig:Scaled_AzVzZ}, where $\tilde{\upsilon}_z$ and $\tilde{z}$ are expressed as:

\begin{subequations}
\label{eq:Scaled_AzVzZ}
\begin{eqnarray}
\label{eq:Scaled_AzVzZ_vz}
\tilde{\upsilon}_z &=& \dfrac{\upsilon_z (t)}{\upsilon_{z0}}\\
\label{eq:Scaled_AzVzZ_z}
\tilde{z} &=& z(t) \cdot \dfrac{1}{\sqrt{M}} \cdot \dfrac{\gamma_\mathrm{D}}{\tan(\beta)}
\end{eqnarray}
\end{subequations}

As shown in Fig.~\ref{fig:Scaled_AzT} and Fig.~\ref{fig:Scaled_AzVzZ}, the quantitative relationships, referring to the Eq.~\eqref{eq:threeparaGamma}, are also valid for the impacting phase, not merely for the peak acceleration and the correlated parameters. Thus, the overlapping trend can be seen as a characteristic curve during a free fall water-entry of a wedge with a wide range of $\upsilon_{z0}$, $\beta$ and $M$ when the instantaneous Froude number $Fr^*$ is greater than 6.5.

\section{Conclusion}
\label{sec:conclusion}

In the present study, the maximal acceleration and correlated parameters of a 2D symmetric wedge \textcolor{black}{free fall water-entry} has been investigated numerically. The effect of deadrise angle and mass on the maximum acceleration, together with several relationships based on the extended von Karman momentum theory, have been thoroughly analyzed. Contributions and findings can be described as:

1)  A significant parameter is the pile-up coefficient $\gamma$, which has been proved dependent on the deadrise angle in the case of the constant velocity water-entry. The dependence law is extended to the \textcolor{black}{free fall water-entry} of the wedge in the present study and the numerical results of $\gamma$ is observed quite close to the similarity solution obtained by Zhao \citep{zhao1993water} by using the original formulation of Dobrovol'skaya \citep{dobrovol1969on}. Furthermore, the maximal acceleration $a_{z\mathrm{max}}$, corresponding time $t^{*}$, velocity $\upsilon_z^{*}$ and penetration depth $z^{*}$ can be predicted theoretically with a good agreement using the extended von Karman momentum theory by combing the equations of original von Karman's momentum theorem and Dobrovol'skaya's solution on $\gamma$.

2)	For the large initial impacting velocity, 5.5 m/s, the maximal vertical acceleration increases when the deadrise angle decreases and it is found herein that the value of maximal vertical acceleration $a_{z\mathrm{max}}$ is proportional to $1/\tan(\beta)$ which is consistent with the theoretical relation. Following the theoretical formulation, the corresponding velocity $\upsilon_z^*$ and the ratio of the corresponding velocity to the initial velocity, $\kappa$, should be independent on the deadrise angle that also can be observed on the numerical results. Looking into other two linear relations, $t^{*}$ and $\upsilon_z^{*}$ with respect to $\tan(\beta)$, they can be established upon the wedge compared with the theoretical relations.
Besides, for the small initial impacting velocity, 1 m/s, due to gravity effect, the theoretical relation is not valid for the ratio $\kappa$ where the value should be constant, 5/6, while the numerical results increase with the deadrise angle. Comparing the data from the two different initial impacting velocity conditions, it is found that the results can be overlapped by increasing the initial velocity, indicating that gravity effect can be reduced or ignored with a larger initial velocity for wedge \textcolor{black}{free fall water-entry}.

3)  With respect to the effect of mass, $a_{z\mathrm{max}}$ as a linear function of $1/\sqrt{M}$ can be observed and the results of $a_{z\mathrm{max}}$ can be further estimated by the combined method with a minor deviation comparing with the numerical results, as well as the theoretical prediction of the corresponding time $t^{*}$ and penetration depth $\upsilon_z^{*}$.
Particularly, gravity effect becomes more significant for the wedge impacting event with a relatively small mass, whereas in the case of large mass the theoretical predictions are no more valid due to the jet root separation.

4)  In order to identify the effective boundary value of the initial impacting velocity, above which the theoretical equations (the combined method) is valid, the instantaneous Froude number, $Fr^*$, is introduced to describe the boundary value. Considering the accuracy below $5\%$ a compromised value, $Fr^* \approx 6.5$, can be used to identify the validity of the combined method.

5)  After investigations, it is believed that the initial velocity $\upsilon_{z0}$, the deadrise angle $\beta$ and the mass $M$ are the three key parameters for the wedge \textcolor{black}{free fall water-entry} condition. Base on these three parameters, a reasonable theoretical estimated approach is established and it can predict the kinematic parameters during impacting phase as well as the variables when the acceleration reaches its peak.


\section*{Acknowledgments}

This work has been supported by China Scholarship Council (CSC, No. 202106830092) and the Project TORPEDO (inTerazione fluidO stRuttura in ProblEmi Di impattO) cooperated in the Institute of Marine Engineering of the National Research Council of Italy. The supports from Open Foundations of EDL Laboratory (EDL19092111), the Aeronautical Science Foundation of China under grant no. 20182352015 and no. F2021110, and Postgraduate Research and Practice Innovation Program of Jiangsu Province (SJCX22-0094) are also acknowledged.








\printcredits

\bibliographystyle{cas-model2-names}

\bibliography{cas-refs}



\end{document}